\documentclass[namedreferences]{solarphysics}

\usepackage[hyperref,optionalrh]{spr-sola-addons} 
\usepackage{graphicx}        
\usepackage{color}           
\usepackage{breakurl}        
\usepackage{rotating,booktabs}

\chardef\us=`\_

\begin{document}

\begin{article}
\begin{opening}

\title{Single ICMEs and Complex Transient Structures in the Solar wind in 2010 -- 2011}

\author[addressref={aff1},corref,email={rodkindg@gmail.com}]{\inits{D.}\fnm{D.}~\lnm{Rodkin}\orcid{http://orcid.org/0000-0002-5874-4737}}
\author[addressref=aff1,email={slem@sci.lebedev.ru}]{\inits{V.}\fnm{V.}~\lnm{Slemzin}\orcid{orcid.org/0000-0002-5634-3024}}
\author[addressref={aff2,aff3},email={andrei.zhukov@sidc.be}]{\inits{A.}\fnm{A.}~\lnm{N. Zhukov}}
\author[addressref=aff1,email={goryaev\underline{ }farid@mail.ru}]{\inits{F.}\fnm{F.}~\lnm{Goryaev}\orcid{orcid.org/0000-0001-9257-4850}}
\author[addressref=aff3,email={jshugai@srd.sinp.msu.ru}]{\inits{Yu.}\fnm{Yu.}~\lnm{Shugay}\orcid{orcid.org/0000-0003-3278-6557}}
\author[addressref={aff3,aff4},email={veselov@dec1.sinp.msu.ru}]{\inits{I.}\fnm{I.}~\lnm{Veselovsky}\orcid{orcid.org/0000-0002-0228-0320}}

\address[id=aff1]{P.N. Lebedev Physical Institute of the Russian Academy of Sciences, 53 Leninskiy Prospekt, Moscow, 119991, Russia}
\address[id=aff2]{Solar-Terrestrial Centre of Excellence -- SIDC, Royal Observatory of Belgium, Ringlaan 3, BE-1180 Brussels, Belgium}
\address[id=aff3]{Skobeltsyn Institute of Nuclear Physics, Lomonosov Moscow State University, Leninskie gory, GSP-1, Moscow, 119991, Russia}
\address[id=aff4]{Space Research Institute (IKI RAS) 84/32, Profsoyznaya str., Moscow, 117810, Russia}

\runningauthor{D.Rodkin \textit{et al.}}
\runningtitle{ICMEs and Complex Transient Structures}

\begin{abstract}

We analyzed statistics, solar sources and properties of interplanetary coronal mass ejections (ICMEs) in the solar wind. In comparison with the first eight years of Cycle~23, during the same period of Cycle~24 the yearly numbers of ICMEs were less correlated with the flare numbers (0.68 vs 0.78) and sunspot numbers (0.66 vs 0.81), whereas the ICME correlation with coronal mass ejections (CMEs) was higher (0.77 vs 0.70). For the period January 2010 -- August 2011, we identified solar sources of the ICMEs included in the Richardson and Cane list. The solar sources of ICME were determined from coronagraph observations of the Earth-directed CMEs supplemented by modeling of their propagation in the heliosphere using the kinematic models (the ballistic and drag-based model) and the Wang-Sheeley-Arge Enlil Cone MHD-based model. A detailed analysis of the ICME solar sources in the period under study showed that in 11 cases out of 23 (48~\%) the observed ICME might be associated with two or more sources. In cases of multiple-source events, the resulting solar wind disturbances may be described as complex (merged) structures occurred due to the stream interactions with properties depending on the type of participating streams. As a reliable marker for identification of interacting streams and their sources, we used the plasma ion composition, as it becomes frozen in the low corona and remains unchanged in the heliosphere. According to the ion composition signatures, we classified these cases into three types: complex ejecta originating from weak and strong CME-CME interactions, as well as merged interaction regions (MIRs) originating from the CME-high-speed stream (HSS) interactions. We described temporal profiles of the ion composition for the single-source and multi-source solar wind structures and compared them with the ICME signatures determined from the kinematic and magnetic field parameters of the solar wind. In single-source events, the ion charge state, as a rule, has one-peak enhancement with average duration of $\sim$~1 day, which is similar to the mean ICME duration of 1.12 days derived from the Richardson and Cane list. In the multi-source events, the total profile of the ion charge state consists of a sequence of enhancements associated with interaction between the participating streams. On average, the total duration of complex structures appearing due to the CME-CME and CME-HSS interactions as determined from their ion composition is 2.4 days, which is more than 2 times longer than that of the single-source events.

\end{abstract}


\end{opening}


\section{Introduction}
\label{S-Introduction}

Interplanetary Coronal Mass Ejections (ICMEs) are the heliospheric counterparts of Coronal Mass Ejections (CMEs). In particular, their special type known as magnetic clouds (MCs), were found to be one of the most geoeffective solar wind transients \citep{burlaga1981,zhang1988,gosling91,farrugia97,zhang2004,echer2005,yermolaev2007,yermolaev12,verbanac13,gopals15}. Various signatures differentiate ICMEs from ambient slow solar wind and flows of fast solar wind from coronal holes \citep{gosling90,zurbuchen06,rich2010}. These signatures include among others, enhanced magnetic field strength, low plasma proton temperatures, bidirectional suprathermal electron strahls (BDEs) and plasma composition anomalies.

Along with solar flares, CMEs are the most powerful events of the large-scale solar activity typically observed by coronagraphs at the distances further than 2~R$_{\odot}$ from the solar center, but the coronagraphic observations give only their density structure and kinematic parameters. Physically, CMEs arise as magnetic eruptions in the low corona, and they are accompanied with different signatures, such as X-ray flares, dimmings, filament disappearances, coronal waves, post-eruptive arcades etc., which may be detected in the EUV wavelength range by regular observations \citep{hudson2001}. In some cases, however, CMEs have no evident signatures on the disk (stealth CMEs: \citealp{robb09}). A comprehensive review of CMEs can be found elsewhere (\textit{e.g.} \citealp{schwenn06,fwebb12}, and references herein).

The rate of CME association with solar flares depends on the solar activity level and flare class. \citet{andrews03} reported that $\sim$~60~\% of M-class flares were associated with CMEs. \citet{yashiro09} examined the CME assoсiations with the flares during Cycle~23 (1996 to 2007) and found that the CME association rate clearly increases with the peak X-ray flux from $\sim$~20~\% for the C-class flares to more than 90~\% for X-class flares.

Among other solar wind parameters, the ion composition data of ICMEs are directly linked with properties of their solar sources because the ion charge states of solar wind transients are frozen-in in the corona, remaining practically unchanged during their propagation in the heliosphere to Earth \citep{hundhausen68}. Thus, the ion composition parameters like $C^{6+}$/$C^{5+}$, $O^{7+}$/$O^{6+}$ ratios and the average iron charge state denoted by $<Q_{Fe}>$ ($Q_{Fe}$ = $\Sigma n_{i}Q_{i}$/$\Sigma n_{i}$, where $n_i$ is the number density of the iron ions with charge state $Q_i$) can be used as tracers to localize the sources from which the transients emerge \citep{hundhausen68,feldman05,hm2016}. \citet{gopals13} presented a statistical relationship between ion composition states of ICMEs and parameters of flares and CMEs during Cycle~23.

\cite{zhang07} investigated the solar and interplanetary sources of 88 geomagnetic storms that occurred during Cycle~23 (1996 -- 2005). The authors identified these sources using as the main signature observations of the halo (full or partial) CMEs and then verifying the surface source region on the front side of the Sun using several eruptive features, including a large scale coronal dimming and a post-eruption loop arcade seen in EUV. They classified the geoeffective events into three categories: (1) S-type, in which the storm is associated with a single ICME and a single CME at the Sun; (2) M-type, in which the storm is associated with a complex solar wind flow produced by multiple interacting ICMEs arising from multiple halo CMEs; (3) C-type, in which the storm is associated with a CIR formed at the leading edge of a high-speed stream originating from a solar coronal hole (CH). They found that S and M-type events constitute 60 and 27~\% of all the events, respectively.

The cases, when a CME interacts with other CMEs, or with a high-speed stream and even with the surrounding coronal magnetic field structures, which can deflect the CMEs from their initial propagation direction or seriously change their kinematic parameters, were considered by \citet{harrison12,lugaz12,temmer12,liu14,liu15,kataoka15,rodkin16,wu16,lugaz17,shugay17}. As a result of this interaction in the heliosphere, large-scale compound structures can reach the Earth in form of complex ejecta when two or more CMEs merge \citep{burlaga1987,burlaga2002}, or merged interaction regions (MIRs, see \citealp{beh1991,burlaga2003,rouillard2010}), if the interaction also involves corotating streams. These structures represent a particular interest for predictive models due to their enhanced geoeffectiveness. Identification and interpretation of such events is more difficult than for non-interacting phenomena because their properties can be modified depending on relations between types and parameters of participating structures. However, ion charge states of the erupted plasma frozen-in in the corona are not modified during interaction in the heliosphere, so they can serve as reliable markers of the solar sources of interacting structures. Analysis of the ion composition measured \textit{in situ} can also be used for validation of the models that account for the interaction.

ICMEs in the beginning and rising phase of Cycle~24 have been studied in a number of recent papers. \citet{kilpua2014} considered solar sources of ICMEs in the minimum between Cycles~23 and 24 (the year 2009).  They found that among 20 ICMEs identified in that period, only seven were seen by \textit{Large Angle and Spectrometric Coronagraph} (LASCO, \citealp{brueckner95}). Eight ICMEs originated from narrow CMEs with the width less than $50^{\circ}$ registered by the coronagraphs onboard the \textit{Sun Earth Connection Coronal and Heliospheric Investigation} (SECCHI) on \textit{Solar Terrestrial Relations Observatory} (STEREO: \citealp{howard08}) but not seen by LASCO. This result demonstrates that occurrence of the full or partial halo CMEs with the width more than $120^{\circ}$ is not a necessary condition for observation of ICMEs arriving to Earth.

\citet{gopals15} studied properties and geoeffectiveness of the special type of ICMEs with enhanced magnetic field strength and smooth rotation of the magnetic field components -- MCs during the first 6 years of Cycles~23 and 24. They noted that although MCs during Cycle~24 appeared more frequently than during Cycle~23, their geoeffectiveness was lower: the mean value of the Dst index in the geomagnetic storms was twice less than during Cycle~23 due to the smaller factor $VBz$ (the product of the MC speed and the out-of-the-ecliptic component of the MC magnetic field). \citet{lawrance16} found that most of geomagnetic storms in the rising phases of Cycles~23 and 24 were associated with ICMEs, the average size of ICMEs during Cycle~23 being larger than during the current Cycle~24. \citet{comp16} investigated some properties of CMEs (speed, acceleration, polar angle, width and mass) using the LASCO data and their association with flares during the period of two solar cycles (1997 -- 2014). They found a linear dependence between logarithms of the flare flux and mass of the corresponding CMEs. They also concluded that the CMEs associated with flares are on average 100~km~s$^{-1}$ faster than the ones not associated with flares. In their work an association of the ICME parameters with properties of their solar sources was not considered. Recently \citet{hess2017} have identified solar sources of 70 Earth-affecting interplanetary coronal mass ejections (ICMEs) during Cycle~24 (2007 -- 2015). The authors analyzed the longitudinal distribution of the sources and found that as in past solar cycles, CMEs from the western hemisphere more likely reached Earth. However, they did not distinguish between single-source and multiple-source events and did not analyze the ICME ion composition.

The aims of this study are to analyze ion charge composition of ICMEs in the rising phase of Cycle~24, to identify their solar sources and to define specific signatures of complex solar wind structures arising as a result of interaction between CMEs and other transient streams like CME and HSS. We analyzed the ion composition parameters $O^{7+}$/$O^{6+}$, $<Q_{Fe}>$ and the $Fe$/$O$ ratio. We present but not analyzed in detail the carbon composition data $C^{6+}$/$C^{5+}$ because of a specific anomaly in carbon ion charge state at high temperatures occurred in the low corona \citep{steiger92,zhao16,kocher2017}. From the observational point of view, Cycle~24 is favorable for investigation of the link between ICMEs and their solar origins. Since May 2010 high-resolution EUV images of the corona taken by the \textit{Atmospheric Imaging Assembly} (AIA: \citealp{lemen12}) telescope onboard the \textit{Solar Dynamic Observatory} (SDO) are available. In 2010 -- 2011 CMEs are observed by coronagraphs onboard the STEREO-A and B spacecraft nearly in quadrature with LASCO onboard the \textit{Solar and Heliospheric Observatory} (SOHO).

\section{Data Sources and Methods}
\label{S-Data}

As the initial data, we used the comprehensive ICME catalog\footnote{http://www.srl.caltech.edu/ACE/ASC/DATA/level3/icmetable2.htm} compiled by Richardson and Cane (hereafter will be referred as the RC list) which is the most complete for Cycle~24 and contains the data for Cycle~23 in the same format, which is convenient for comparison. We took 1-hour averaged values  of the proton speed, density, temperature and magnetic field components from the Level~2 data of the \textit{Solar Wind Electron Proton Alpha Monitor} (SWEPAM: \citealp{mccomas98}) and magnetometer (MAG: \citealp{smith1998}) onboard the \textit{Advanced Composition Explorer} (ACE: \citealp{stone98}). In the analysis of the ICME ion composition, we used 1-hour averaged data from the \textit{Solar Wind Ion Composition Spectrometer} (SWICS: \citealp{gloecker98}) for the period before 23 August 2011 (SWICS~1.1 data\footnote{http://www.srl.caltech.edu/ACE/ASC/level2/index.html}). In the periods of the ACE data gaps we used also the data from \textit{Solar Wind Experiment} onboard WIND (SWE: \citealp{ogil1995}). Due to recalibration of the SWICS instrument, the ion composition data after 23 August 2011 are not fully compatible with the older data, so we limited our investigation to the period before this date. To identify the Earth directed CMEs, we used SDO/AIA data\footnote{https://sdo.gsfc.nasa.gov/} \citet{lemen12},in addition to the LASCO CME catalogs provided by the Cordinated Data Analysis Workshops (CDAW)\footnote{https://cdaw.gsfc.nasa.gov/}, Solar Eruptive Event Detection System (SEEDS)\footnote{http://spaceweather.gmu.edu/seeds/secchi.php} and Computer Aided CME Tracking (CACTus)\footnote{https://secchi.nrl.navy.mil/cactus/} databases.

\subsection{Statistics of ICMEs and Associations with Solar Activity during the first eight years of Cycles~23, 24}
\label{S-Statistics and associations}

Figure~1 shows the yearly total numbers of ICMEs, MCs from the RC list, X-ray flares from the GOES database, CMEs from CDAW database during Cycles~23 (1996 -- 2008), 24 (2009 -- 2016) in comparison with variation of the solar activity (the yearly total sunspot numbers from the Sunspot Index and Long-term Solar Observations database of the Solar Influences Data Analysis Center SIDC/SILSO\footnote{http://sidc.oma.be/silso/}). Table~1 shows the total numbers of flares, CMEs, ICMEs, MCs and sunspot numbers (SN) during the first eight years of Cycles~23, 24 and correlations between their yearly frequencies. During the period 2009 -- 2016 (Cycle~24) the total number of ICMEs was 29~\% less than during the period 1996 -- 2003 (Cycle~23), which correlates with the decrease of the total sunspot numbers. The fraction of MCs among all ICMEs during Cycle~24 is larger than during the previous Cycle~23 (0.79 vs 0.62) for the first eight years of these cycles.

\begin{figure}[tbp] 
\centerline{
\includegraphics[width=0.55\textwidth,clip=]{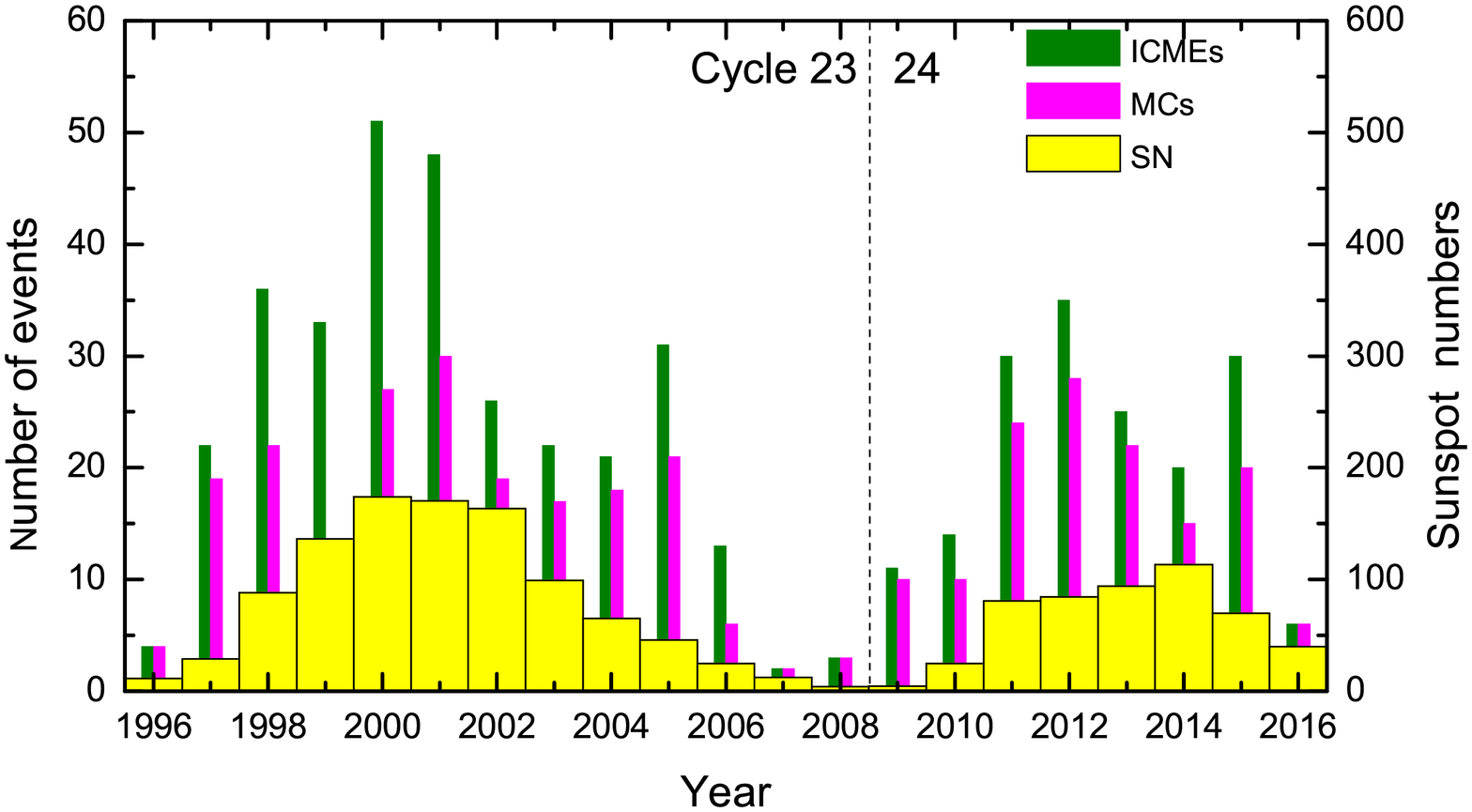}
\hspace*{-0.08\textwidth}
\includegraphics[width=0.55\textwidth,clip=]{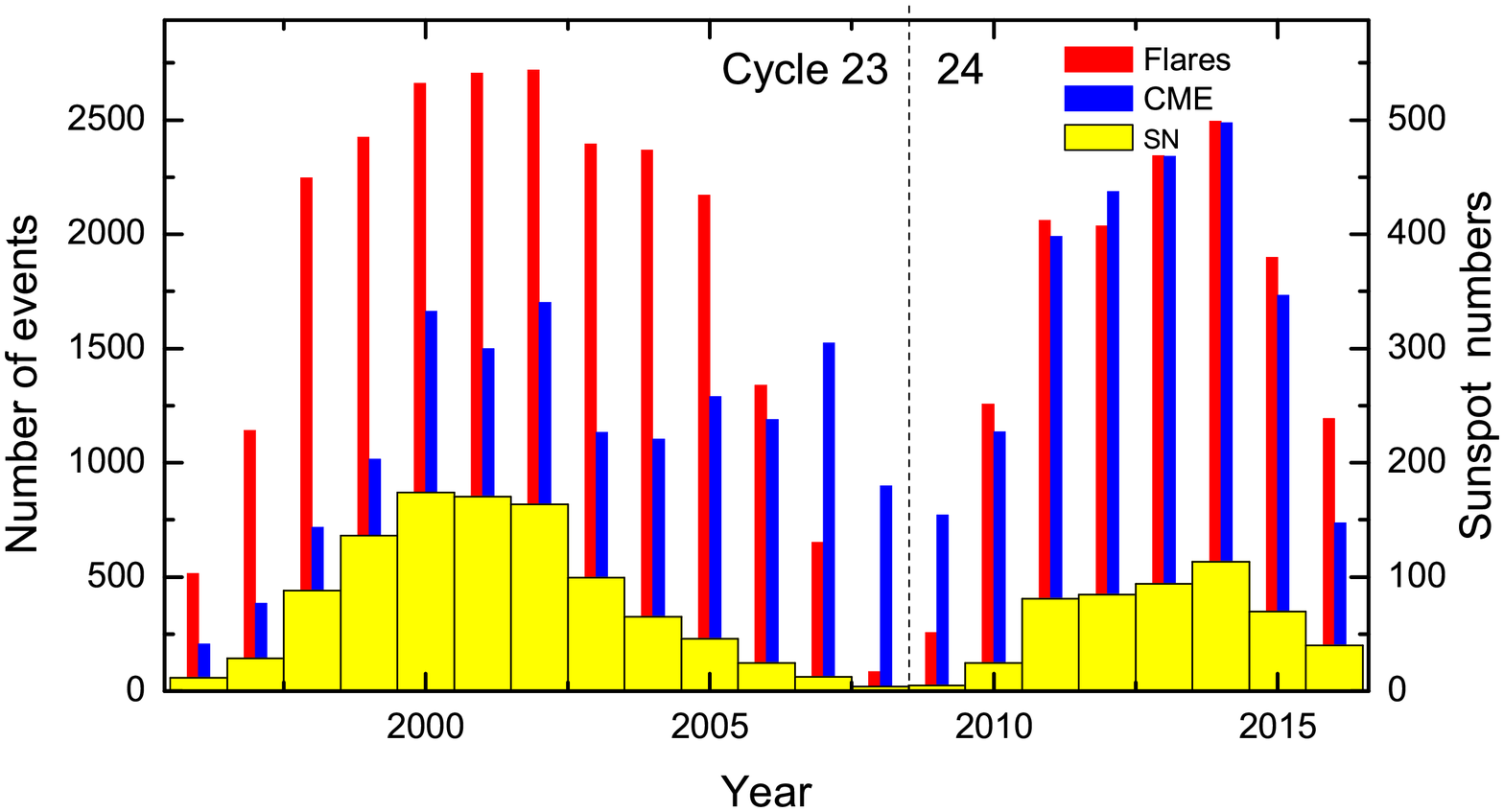}
}
\caption{Left panel: Yearly total numbers of ICMEs (green bars), MCs (magenta bars)
and sunspot numbers (yellow bars) during Cycles~23 and 24 (1996 -- 2016). Right panel: Yearly
total numbers of CMEs (blue bars), X-ray flares (red bars) and sunspot numbers for the same
period.}
\label{fig1}
\end{figure}

\begin{table}[tbp]  
\caption{Total numbers of flares, CMEs, ICMEs, MCs and sunspot numbers (SN) during the first eight years of Cycles~23, 24 and correlations between their yearly frequencies }
\label{Total numbers}
\begin{tabular}{ccc}     
\hline                   
Solar cycle & 23 (1996-2003) & 24 (2009-2016) \\
\hline
Total number of & 16832 & 13542 \\
flares (GOES) & & \\
\hline
Total number of & 8321 & 13385 \\
CMEs (LASCO) & & \\
\hline
Total number of ICMEs (RC) & 242 & 171 \\
\hline
Portion of MCs (RC) & 0.62 & 0.79 \\
\hline
Correlations of the & 0.70 & 0.77 \\
year numbers ICME-CME & & \\
\hline
Correlations of the & 0.78 & 0.68 \\
year numbers ICME-flares & & \\
\hline
Correlations of the & 0.81 & 0.66 \\
year numbers ICME-SNs & & \\
\hline
Correlations of the & 0.96 & 0.95 \\
year numbers CME-SNs & & \\
\hline

\end{tabular}
\end{table}

The total number of CMEs during the first eight years of Cycle~24 was 61~\% larger than in the same period of Cycle~23, but the number of X-ray flares was 20~\% smaller. In the beginning of Cycle~24 the number of the X-ray flares registered by GOES was two times less than in Cycle~23 because most of them were one or two orders weaker than in the previous cycle, and in many cases X-ray fluxes were below the instrumental sensitivity threshold. In 2009, the X-ray photometer SPHINX onboard the CORONAS-Photon/TESIS telescope, having the sensitivity level two orders below than GOES, detected 963 flares whereas GOES detected only 256 flares \citep{mrozek13}. The correlations between the yearly frequencies of ICMEs and CMEs for the first eight years of Cycles~23 and 24 are comparable -- 0.70 and 0.77, whereas the correlation between the ICME and the X-ray flare yearly numbers amounts to 0.78 for Cycle~23 and 0.68 for Cycle~24. At the same time, the strength of X-ray flares in Cycle~24 is noticeably less than during the previous one. Similarly, many weak CMEs and CME-like events were observed during Cycle~24 \citep{kilpua2014}.

Solar flares and CMEs result from the rapid energy dissipation in the solar atmosphere. They can proceed jointly with different partitions of energy between them from one event to another \citep{emslie2012}. The energy partition ratio $\alpha = E_{rad}/E_{tot}$ of the emitted electromagnetic energy $E_{rad}$ to the total emitted energy $E_{tot} = E_{rad} + E_{pl}$ in different cases may vary from one to zero, with $E_{pl}$ being the kinetic energy of the ejected plasma (CME). The limiting case $\alpha=0$ corresponds to a pure CME without any flare. The limiting case $\alpha=1$ corresponds to a pure flare without a CME, which is called a confined flare. Intermediate cases $0< \alpha <1$ correspond to eruptive flares. Our analysis shows a difference between Cycles~23 and 24 in this regard. As it follows from Figure~1, the confined flares are relatively more frequent phenomena during Cycle~23 because the closed magnetic fields associated with a larger number of sunspots was stronger. One can speculate that CMEs and flares correlate better during the weaker 24th cycle because of the same reason weaker closed magnetic fields in active regions.

\subsection{Identification of ICME Solar Sources for the period from January 2010 to August 2011}
\label{S-Solar sources and composition}

To determine relations between parameters of ICMEs and their solar sources, we first need to establish associations between the ICMEs and the solar activity events. We considered the temporal boundaries of the ICMEs taken from the RC list which were defined by such signatures as enhanced velocity of protons $V_p$ and magnetic field strength $|B|$, decreased temperature of protons $T_p$ below the expected from the $V_p$ value. The ion composition data are mentioned as one of the main signatures of ICMEs in the paper by \citet{rich04}. However, due to the corresponding data being not shown in the RC catalog, and the temporal boundaries of the ICMEs corresponding to the plasma temperature and magnetic field parameters rather than the ion charge state enhancements, the composition data were probably not considered as the main signature in the routine identification of the ICME boundaries.

The first step was to determine the time intervals at the Sun when one or several CMEs, in principle, can produce the given ICME. We calculated these time intervals using the average ICME velocity in the ballistic approximation \citep{nolte73a} with an accuracy of $\sim$~12~h \citep{mcneice11}.

The second step was to select CMEs directed to Earth. For this purpose, we used the data from the coronagraphs COR2 at the STEREO-A and B spacecraft. During 2010 -- 2011 both STEREO spacecraft were positioned nearly in quadrature with the direction to Earth with deviation being less then $\pm~40^{\circ}$. Firstly, we selected the CMEs, if they appearing in STEREO-A COR2 at the East limb and in STEREO-B COR2 at the West limb within the time interval of 1 hour, with the latitudinal contours crossing the equatorial plane. The CME velocities determined by the COR2 on STEREO-A and B in all cases, except two fastest CMEs, differ less than on 30~\%, which, taking into account positions of the STEREO spacecraft, correspond to the CME propagation angles with respect to the Sun-Earth line of less than 20$^{\circ}$. Then we analyzed also the LASCO C2 data, looking in particular for full and partial halo CMEs. Among 24 CMEs registered by the STEREO coronagraphs, 16 were observed by LASCO, from which 13 were halo or partial halo type. For example, for the two fastest cases mentioned above LASCO data show halo CMEs, so we still consider these CMEs as directed at the Earth even if significant differences can be seen in the speeds measured by COR2~A and COR2~B. As an additional criterion of association of a CME with the given ICME, the time of arrival of CME to Earth calculated with the Drag-based prediction model \citep{zic15} should lie within the temporal limits of the ICME $\pm$~12~hours \citep{shi15}.

The associated X-ray flares were determined under the condition that the time interval between the flare peak and the start-up time of the selected CME did not exceed 1~hour. Spatial associations between flares and CMEs were established using observations of coronal dimmings, erupting filaments, and post-eruption arcades (see \textit{e.g.} \citealp{hudson2001,zhukov2007}) in the EUV imaging data from SOHO/EIT (EIT: \citealp{delab1995}) (for the period from January to May 2010) and SDO/AIA (for the period from May 2010 to August 2011). In this way, the RC list was supplemented with the data on the identified solar sources (CMEs and flares) for the ascending phase of Cycle~24 (January 2010 -- August 2011).

In total, the list for the period from January 2010 to August 2011 contains 23 events (see Table~2 and Table~3). 12 of them were linked with the single-source (SS) CMEs and 11 are the multi-source (MS) events (Fig.~2). Only in 10 cases LASCO observed full or partial halo CMEs. In one more case, a narrow (angular width $53^{\circ}$) non-halo CME was detected by LASCO. In the remaining 12 cases, LASCO did not detect a CME, although STEREO observations nearly in quadrature with LASCO indicate that CMEs did occur in 6 of these events. This means that the detection of a full or a partial halo CME by a remote-sensing observatory at 1~AU (\textit{e.g.} SOHO) is not a necessary condition for a subsequent ICME detection \textit{in situ} around the same observatory (\textit{e.g.} by SOHO or ACE).

\begin{table}[tbp]  
\caption{The ICMEs from RC list detected in 2010 and parameters of their possible solar sources. }
\label{ICME1}
\begin{tabular}{cccccccc}     
\hline                   
 & ICME (RC) & Dst & STA/STB & V$_A$/V$_B$, & LASCO & V$_{LASCO}$, & Flare (GOES) \\
 & start/end (UT)& nT & Time, UT & km~s$^{-1}$ & Time, UT& km~s$^{-1}$ & Class and \\
 & & & & & and type & & peak time, UT\\
\hline
1 & 1 Jan. 22:00/ & -4& N/A & N/A & N/A & N/A & N/A \\
& 3 Jan. 10:00 & & & &  & &  \\
\hline
2 & 7 Feb. 18:00/ &-22 & 3 Feb. & 156/177 &N/A & N/A& N/A\\
& 8 Feb. 22:00 & & 3:24/3:54 (C)$^*$ & & & & \\
\hline
3 & 11 Feb. 8:00/ &-7  & 6 Feb. & 349/284 & 6 Feb. & 240 & M2.9, 6 Feb. \\
& 12 Feb. 3:00 & & 19:54/20:54 (C) & & 20:06 & & 18:59 \\
& & & 7 Feb.  & 528/466 & 7 Feb.  & 420 & C1.1, 7 Feb. \\
& & & 4:24/3:54 (S) & & 3:54 (H)$^{**}$ & & 3:29 \\
& & & 6 Feb.  & 325/310 & 6 Feb.  & 240 &C4.0, 6 Feb.  \\
& & & 7:54/9:54 (S) & & 8:06 & & 7:04 \\
\hline
& 19 Feb. 15:00/  &-11 & & & & & \\
4 & 20 Feb. 18:00 & & 16 Feb.  & 543/568& N/A& N/A& N/A\\
& 21 Feb. 0:00/ &-6 & 6:54/7:24 (C)& & & & \\
& 22 Feb. 0:00 & & & & & & \\
\hline
5 & 22 Feb. 13:00/  &-16 & N/A& N/A& N/A&N/A &N/A \\
& 22 Feb. 22:00 & & & & & & \\
\hline
6& 5 Apr. 12:00/  &-81 & 3 Apr.  & 833/833 & 3 Apr.  & 670 &B7.4, 3 Apr.  \\
& 6 Apr. 14:00 & & 9:54/10:24 (C) & & 10:30 (H) & & 9:54 \\
\hline
7 & 9 Apr. 18:00/  &-31 & 6 Apr. & 807/781& N/A& N/A& N/A\\
& 10 Apr. 16:00 & & 1:54/1:54 (C) & & & & \\
\hline
8 & 12 Apr. 1:00/  &-67 & 8 Apr.  & 520/540& 8 Apr. & 264&B3.7, 8 Apr. \\
& 12 Apr. 15:00 & & 3:54/4:24 (C) & & 4:54 (PH) & & 3:25 \\
\hline
9 & 30 Apr. 6:00/  &-13 & 26 Apr. & 480/347&N/A & N/A& N/A\\
& 1 May 12:00 & & 12:54/13:24 (C) & & & &  \\
\hline
10 & 28 May 19:00/  &-80 & 23 May & 363/365& 23 May & 258&B1.3, 23 May  \\
& 29 May 17:00 & & 17:54/17:54 (S) & & 18:06 (H) & & 18:01 \\
& & & 24 May & 504/536& 24 May & 427&B1.1, 24 May \\
& & & 14:24/14:54 (S)& & 14:06 (H) & & 14:06\\
\hline
11 & 21 June 6:00/  &-11 & N/A& N/A& N/A& N/A& N/A\\
& 22 June 14:00 & & & & & & \\
\hline
12 & 4 Aug. 10:00/  &-74 & 1 Aug & 670/528& N/A& N/A& N/A\\
& 5 Aug. 0:00 & & 3:54/4:24 (S) & & & & \\
& & & 1 Aug. & 1100/906& N/A& N/A&C32, 1 Aug. \\
& & & 8:24/8:54 (S)&& & & 8:26\\
\hline
13 & 28 Dec. 3:00/  &-43 & 23 Dec.  &320/337 & 23 Dec. &286 & N/A\\
& 28 Dec. 15:00 & & 5:54/6:54 (C) & & 5:00 (PH) & &\\
\hline
& & & & & & & \\
\multicolumn{8}{l}{$^{*}$C -- CACTus data, S -- SEEDS data.}\\
\multicolumn{8}{l}{$^{**}$H -- Halo CME, PH -- Partial Halo CME.}\\
\end{tabular}
\end{table}

\begin{table}[tbp]  
\caption{The ICMEs from RC list detected in 2011 and parameters of their possible solar sources.}
\label{ICME2}
\begin{tabular}{cccccccc}     
\hline                   
 & ICME (RC) & Dst & STA/STB & V$_A$/V$_B$, & LASCO & V$_{LASCO}$, & Flare (GOES)\\
 & start/end (UT)& nT & Time, UT & km~s$^{-1}$ & Time, UT& km~s$^{-1}$ & Class and \\
 & & & & & and type & & peak time, UT\\
\hline
14 & 24 Jan. 10:00/ &-14 &N/A &N/A& N/A&N/A& N/A\\
& 25 Jan. 12:00 & & & & & &\\
\hline
15 & 4 Feb. 13:00/ &-63 & 30 Jan.  &219/235 & 30 Jan. & 120 & N/A\\
& 4 Feb. 20:00 & & 3:54/9:24$^*$ (C) & & 12:36 (PH) & &\\
\hline
16 & 18 Feb. 19:00/ &-32 & 14 Feb. &349/386 & 14 Feb. 2011 &326 &C6.6, 14 Feb. \\
& 20 Feb. 8:00 & & 19:24/19:54 (S) & & 18:24 (H) & & 19:30\\
& & & 15 Feb. &844/834 & 15 Feb. &669 &X2.2, 15 Feb. \\
& & & 2:24/2:24 (S) & & 2:24 (H) & & 1:56\\
\hline
17 & 6 Mar. 9:00/ &-27 & 3 Mar.  &211/261 & 3 Mar.  &263 & N/A\\
& 8 Mar. 6:00 & & 3:54/4:54 (S) & & 6:12 (PH) & &\\
& & & 4 Mar.  &340/350 & N/A&N/A & N/A\\
& & & 0:54/0:54 (S) & & & &\\
\hline
18 & 29 Mar. 23:00/  &-4 & N/A&N/A& N/A&N/A & N/A\\
& 31 Mar. 4:00 & & & & & &\\
\hline
19 & 28 May 5:00/ &-80 & 24 May &640/657 & N/A &N/A & N/A\\
& 28 May 21:00 & & 15:54/16:24 (C) & & & &\\
\hline
20 & 6 July 17:00/  &-28 & N/A &N/A & N/A &N/A & N/A\\
& 7 July 12:00 & & & & & &\\
\hline
21 & 15 July 4:00/ &-17 & 11 July  &446/499 & 11 July  &266 &C2.6, 11 July\\
& 16 July 15:00 & & 11:24/11:24 (C) & & 12:00 & & 11:03\\
\hline
22 & 5 Aug. 5:00/  &-15 & 2 Aug. &780/1029 & 2 Aug. &712 &M1.4, 2 Aug. \\
& 5 Aug. 14:00 & & 5:54~(C)/6:54~(S) & & 6:36 (PH) & & 6:19\\
\hline
23 & 6 Aug. 22:00/  &-115 & 3 Aug. & 892/833 & 3 Aug.  & 610 &M6.0, 3 Aug.\\
& 7 Aug. 22:00 & & 13:54/13:54 (c) & & 14:00 (H) & & 13:48\\
& & & 4 Aug.  & 1193/1562 & 4 Aug. & 1315 &M9.3, 4 Aug.\\
& & & 3:54/4:24 (c) & & 4:12 (H) & & 3:57\\
\hline
& & & & & & & \\
\multicolumn{8}{l}{$^{*}$This time interval is more than 1 hour due to very slow poor ejecta (see Section~3.2.3).}\\

\end{tabular}
\end{table}

\begin{figure}[tbp] 
\centerline{
\includegraphics[width=0.55\textwidth,clip=]{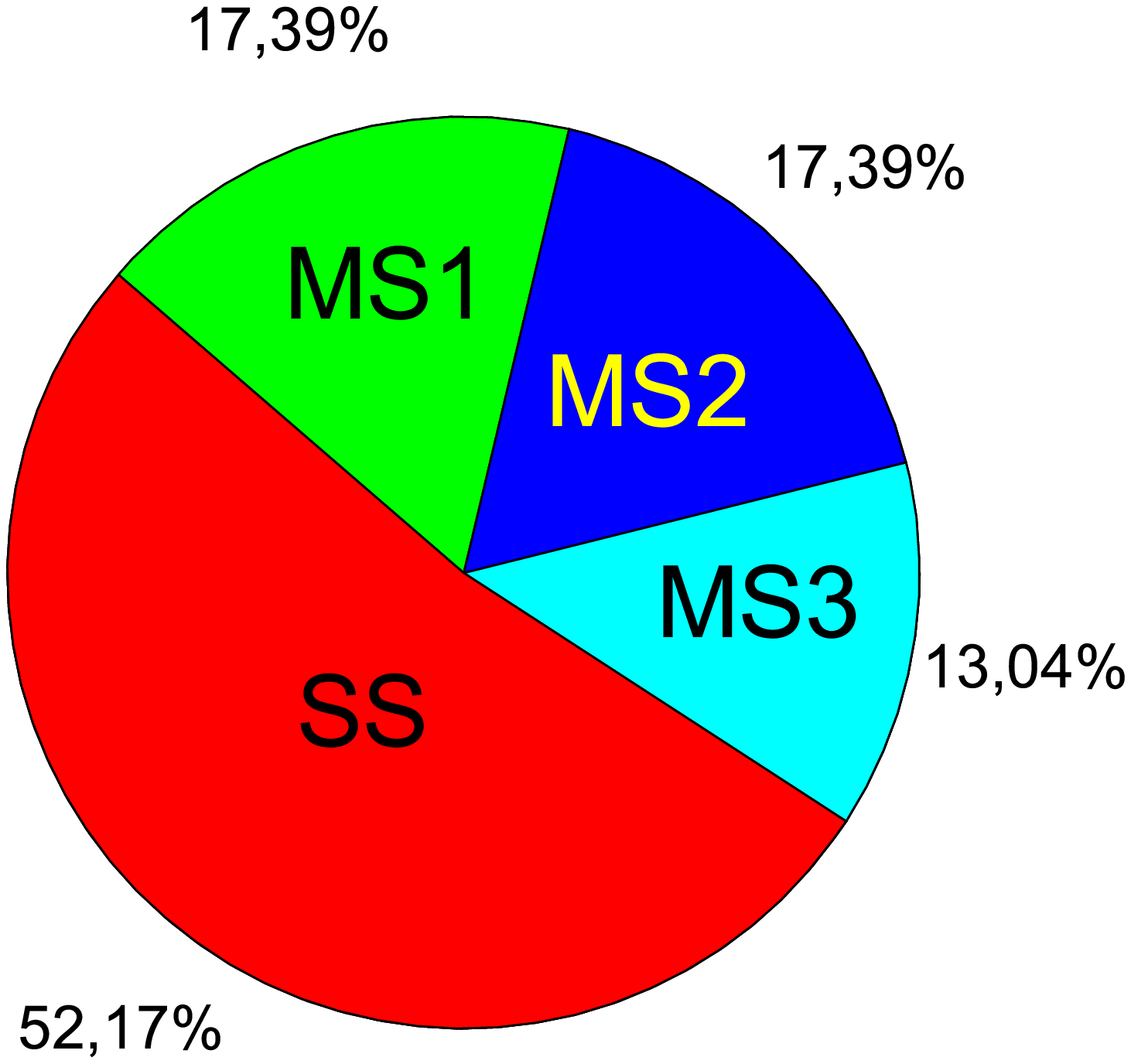}
\hspace*{-0.08\textwidth}
\includegraphics[width=0.55\textwidth,clip=]{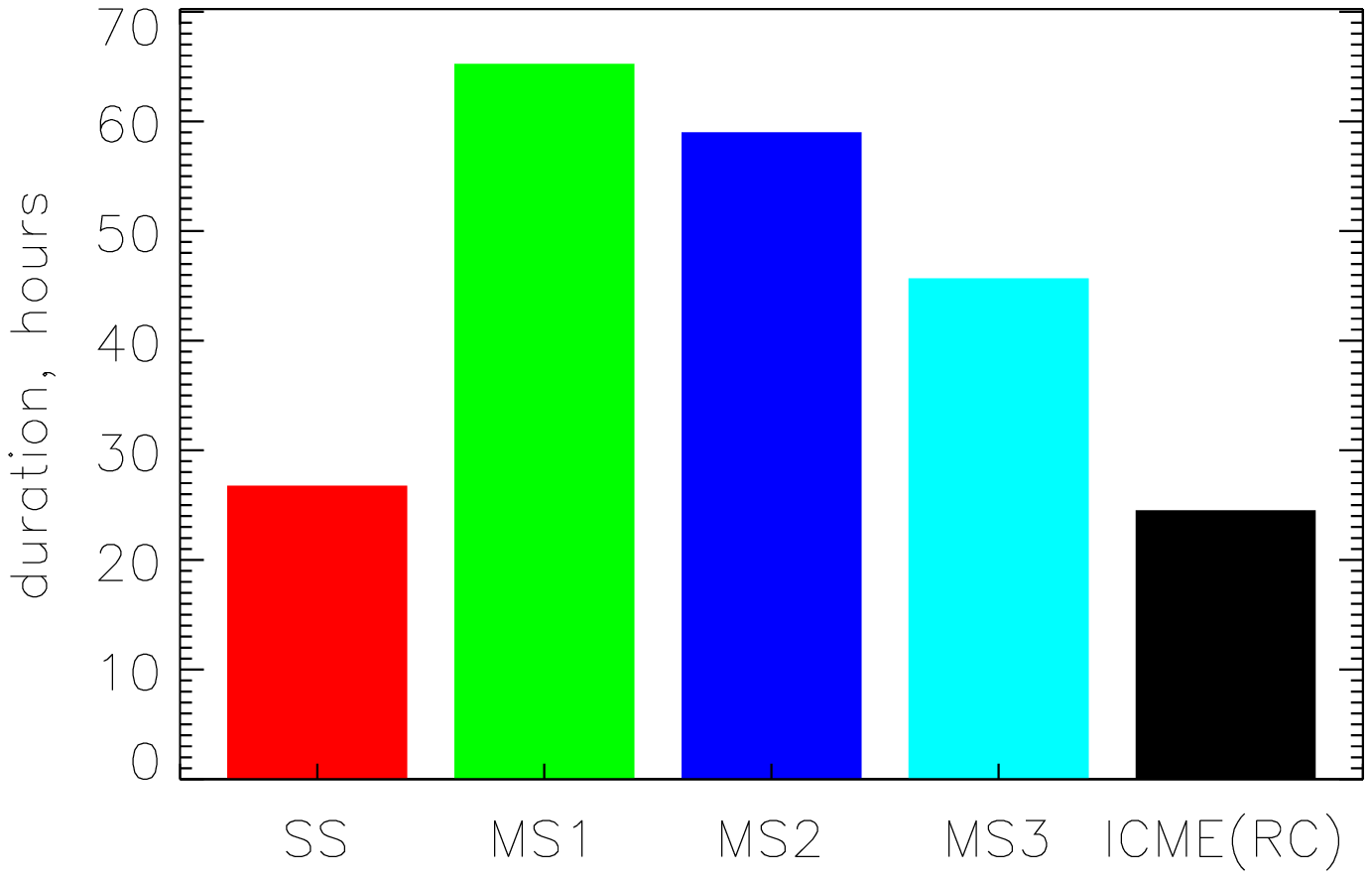}
}
\caption{ Left panel: Portions of different types of events observed during January 2010 -- August 2011. SS -- single-source; MS1 -- complex ejecta with weak interaction; MS2 -- complex ejecta with strong interaction; MS3 -- complex structure originated from CME-HSS interaction. Right panel: average durations of these events and ICMEs (RC catalog) during January 2010 -- August 2011.}
\label{fig2}
\end{figure}

Besides CMEs, we must take into account other types of solar wind as sources, such as corotating interaction regions (CIRs) and heliospheric current sheet (HCS), which can interact with CMEs and change the parameters of the events observed \textit{in situ} at 1~AU. Below we will follow the definition of \citet{burlaga2003} and name such complex events as merged interaction regions (MIRs).

\subsection{Ion Composition Parameters of ICMEs for the period from January 2010 to August 2011}
\label{S-CompParameters}

The averaged values of $<Q_{Fe}>$ and the $O^{7+}$/$O^{6+}$ ratio for the ICMEs in the period 2010 -- 2011 are lower than those during Cycle~23 due to weaker heating processes in the corona (Table~4). Maximum values of $<Q_{Fe}>$ were approximately the same for both cycles. However, for $O^{7+}$/$O^{6+}$ ratio the maximum values were higher during 23rd Cycle.

\begin{table}[tbp]  
\caption{Comparison of maximum/mean values of the ion composition parameters averaged over the ICME duration for the rising phase periods of 23rd \citep{gopals13} and 24th solar cycles (calculated for the 2010--2011 period under study).}
\label{comparison}
\begin{tabular}{cccc}     
\hline                   
Solar cycle & & $<Q_{Fe}>$ & $O^{7+}$/$O^{6+}$ \\
\hline
23 & Max & 10.6 -- 17.7 & 0.3 -- 5.3 \\
(1998 -- 2000)& Avg & 11.9 & 0.7 \\
\hline
24 & Max & 10.2 -- 17.6 & 0.2 -- 1.5 \\
(2010 -- Aug. 2011)& Avg & 10.8 & 0.34 \\
\hline

\end{tabular}
\end{table}

The ion composition parameters of the ICMEs, such as the maximal, minimal and averaged values of the temperature-dependent ratios $O^{7+}$/$O^{6+}$, the average charge of the iron ions $<Q_{Fe}>$ and the magnetic structure-dependent ratio $Fe$/$O$, were determined from the ACE data for SS and MS events appeared in the period 2010 -- August 2011 (see Table~5 and Table~6). In these tables, we use the notation "ICME" for the solar wind disturbances identified in RC list. The notation "Event" we use in the case of SS and MS events identified with the help of ion composition analysis. The start and end times of ICMEs were taken from the RC list. Intervals of SS and MS events we defined from the enhancements of the solar wind ion charge states. Below we consider examples of the SS and MS events identified from the ACE solar wind data in the period from January 2010 to August 2011.

\begin{sidewaystable} 
\setlength\tabcolsep{4pt} 
\footnotesize  
\centering
\caption{ Parameters of single- and multi-source transient events in the solar wind during the year 2010. Periods of events were defined from the enhancements of the solar wind ion charge state. SS -- single-source; MS1 -- complex ejecta with weak interaction; MS2 -- complex ejecta with strong interaction; MS3 -- complex structure originated from CME-HSS interaction. In the last column, the duration of the event is compared with the duration of the ICME mentioned in the RC list.}
\centering
\label{param1}
\begin{tabular}{ @{} p{0.5cm} *{17}{cc} @{}}
\hline                   
&Type & Event& $n_p$,  &$V_p$, & $T_p$, & \multicolumn{3}{c}{$O^{7+}$/$O^{6+}$} & \multicolumn{3}{c}{$Fe$/$O$}& \multicolumn{3}{c}{$Q_{Fe}$}& $avg |B|$, &ICME/Event\\
& & start/end (UT) & $1/cm^{3}$ & km~s$^{-1}$ & 10$^4$ K & max & avg & min & max & avg & min & max & avg & min & nT & duration, h\\
\hline
1&SS&1 Jan. 2010 22:00/ & 8.9 &284 &2.45 &0.15 &0.09&0.03&0.62&0.21&0.03&10.5&9.3&8.3&6.8& 36/36\\
&&3 Jan. 2010 10:00 & & & & &&&&&&&&&&\\
\hline
2&SS&7 Feb. 2010 18:00/ & 7.3 &364 &2.08 &0.39 &0.17&0.07&0.47&0.27&0.14&10.7&9.8&9&8.7&28/28\\
&&8 Feb. 2010 22:00 & & & & &&&&&&&&&&\\
\hline
3&MS1&11 Feb. 2010 0:00/ & 5.7 &346 &2.61 &1.02 &0.42&0.06&0.46&0.2&0.1&13.8&10.9&9&6.5&19/59\\
&&13 Feb. 2010 11:00 & & & & &&&&&&&&&&\\
\hline
4&MS1&18 Feb. 2010 12:00 &2.9 &418 &6.55 &0.35 &0.14&0.03&0.65&0.19&0.02&12&10&7.6&6.7&51/86\\
&&22 Feb. 2010 2:00 & &&&&&&&&&&&&&\\
\hline
5&SS&22 Feb. 2010 8:00/ & 5.7 &360 &3.22 &0.31 &0.15&0.05&0.16&0.1&0.05&10.7&9.5&8.6&5.6&9/16\\
&&23 Feb. 2010 0:00 &  & & & &&&&&&&&&&\\
\hline
6&SS&5 Apr. 2010 12:00/ & 2.2 &631 &5.12 &1.09 &0.54&0.06&0.29&0.19&0.1&13.7&12.4&10&9.2&26/26\\
&&6 Apr. 2010 14:00 &  & & & &&&&&&&&&&\\
\hline
7&SS&9 Apr. 2010 0:00/ & 2.9 &419 &3.17 &0.41 &0.18&0.07&0.8&0.21&0.09&11.5&10.3&9.1&3.5&22/59\\
&&11 Apr. 2010 11:00 &  & & & &&&&&&&&&&\\
\hline
8&SS&12 Apr. 2010 0:00/ & 11.2 &408 &2.05 &1.17 &0.42&0.19&0.22&0.18&0.12&14.7&11.9&10.2&10.1&14/16\\
&&12 Apr. 2010 16:00 &  & & & &&&&&&&&&&\\
\hline
9&MS3&30 Apr. 2010 6:00/ & 6.1 &405 &7.47 &0.36 &0.16&0.01&0.74&0.2&0.03&11.7&9.6&7&6.0&30/66\\
&&3 May 2010 0:00 &  & & & &&&&&&&&&&\\
\hline
10&MS2&28 May 2010 2:00/ & 10.8 &392 &7.28 &1.11 &0.34&0.03&0.4&0.16&0.06&12.5&10.1&9.1&10.1&22/69\\
&&30 May 2010 23:00 &  & & & &&&&&&&&&&\\
\hline
11&SS&21 June 2010 7:00/ & 5.8 &359 &2.35 &0.31 &0.21&0.14&0.39&0.23&0.11&10.3&9.8&9.2&6.1&32/31\\
&&22 June 2010 14:00 &  & & & &&&&&&&&&&\\
\hline
12&MS2&3 Aug. 2010 18:00/ & 7.3 &536 &9.77 &0.6 &0.26&0.07&0.57&0.16&0.04&12.4&10.3&9.2&9.5&14/38\\
&&5 Aug. 2010 8:00 &  & & & &&&&&&&&&&\\
\hline
13&SS&28 Dec. 2010 3:00/ & 16.5 &347 & 3.43 &0.54 &0.28&0.08&0.22&0.1&0.05&12.9&10.1&9.1&9.2&12/15\\
&&28 Dec. 2010 18:00 &  & & & &&&&&&&&&&\\
\hline

\end{tabular}
\end{sidewaystable}

\begin{sidewaystable} 
\setlength\tabcolsep{4pt} 
\footnotesize  
\centering
\caption{ Parameters of single- and multi-source transient events in the solar wind during the year 2010. Periods of events were defined from the enhancements of the solar wind ion charge state. In the last column, the duration of the event is compared with the duration of the ICME mentioned in the RC list.}
\centering
\label{param2}
\begin{tabular}{ @{} p{0.5cm} *{17}{cc} @{}}
\hline                   
&Type & Event& $n_p$  &$V_p$, & $T_p$, & \multicolumn{3}{c}{$O^{7+}$/$O^{6+}$} & \multicolumn{3}{c}{$Fe$/$O$} & \multicolumn{3}{c}{$Q_{Fe}$}& $avg |B|$, &ICME/Event\\
&& start/end (UT)& $1/cm^{3}$ &  km~s$^{-1}$ & 10$^4$ K & max & avg &min & max & avg&min & max & avg&min & nT &duration, h\\
\hline
14&SS&24 Jan. 2011 10:00/ & 7.6 &350 &1.74 &0.46 &0.27&0.09&0.17&0.1&0.05&11.9&10.6&9&6.6&28/28\\
&&25 Jan. 2011 12:00 &  & & & &&&&&&&&&&\\
\hline
15&MS3&4 Feb. 2011 1:00/ & 12.0 &452 &16.23 &0.47 &0.23&0.01&1.14&0.29&0.07&12.2&10.2&9.4&9.7&7/28\\
&&5 Feb. 2011 5:00 &  & & & &&&&&&&&&&\\
\hline
16&MS2&18 Feb. 2011 0:00/ & 4.2 &471 &11.98 &1.98 &0.94&0.15&0.83&0.12&0.02&17.6&13.8&9.7&10.5&37/68\\
&&20 Feb. 2011 20:00 &  & & & &&&&&&&&&&\\
\hline
17&MS1&6 Mar. 2011 2:00/ & 5.3 &390 &3.27 &0.6 &0.29&0.03&0.28&0.14&0.01&12.6&9.8&8.4&5.2&45/74\\
&&10 Mar. 2011 0:00 &  & & & &&&&&&&&&&\\
\hline
18&MS1&30 Mar. 2011 0:00/ & 2.8 &349 &1.8 &1.23 &0.45&0.16&1.06&0.44&0.1&11.8&10.4&9.6&11.8&29/42\\
&&31 Mar. 2011 18:00 &  & & & &&&&&&&&&&\\
\hline
19&MS3&28 May 2011 5:00/ & 3.4 &617 &26.23 &0.79 &0.17&0.01&0.3&0.16&0.09&13.6&10.3&9.1&8.5&16/43\\
&&30 May 2011 0:00 &  & & & &&&&&&&&&&\\
\hline
20&SS&6 July 2011 17:00/ & 5.7 &359 &2.84 &0.52 &0.26&0.09&0.86&0.37&0.06&11.5&9.5&8.7&5.0&19/28\\
&&7 July 2011 21:00 &  & & & &&&&&&&&&&\\
\hline
21&SS&15 July 2011 4:00/ & 5.8 &420 &2.00 &1.08 &0.53&0.18&0.21&0.14&0.07&15.2&13.4&9.2&4.3&35/26\\
&&16 July 2011 6:00 &  & & & &&&&&&&&&&\\
\hline
22&SS&5 Aug. 2011 4:00/ & 3.2 &416 &5.67 &0.36 &0.24&0.11&0.55&0.3&0.13&11.5&10.1&9.3&4.8&9/12\\
&&5 Aug. 2011 16:00 &  & & & &&&&&&&&&&\\
\hline
23&MS2&5 Aug. 2011 17:00/ & 4.0 &537 &16.68 &1.32 &0.34& 0.03&0.51&0.23&0.07&15.9&11.3&9.5&8.4&24/61\\
&&8 Aug. 2011 6:00 &  & & & &&&&&&&&&&\\
\hline

\end{tabular}
\end{sidewaystable}

\section{Single- and Multiple-source events in the period from January 2010 to August 2011}
\label{S and M source ICMEs}

We have analyzed 23 events occurred in the period January 2010 -- August 2011 and found their possible solar sources. We distinguish the single-source events (having one probable source) and multiple-source events (having more than one solar source). The typical parameters of the single-source ICMEs in comparison with that of CIRs and ambient solar wind established during Cycle~23 \citep{zurbuch1999,lepri2001,cane2003,borovsky2006,galvin2009,rich2010,mason2012,gopals13} are given in Table~7.

\begin{table}[tbp]  
\caption{Typical parameters of solar wind streams for 23rd solar cycle.}
\label{typical parameters}
\begin{tabular}{ccccccc}     
\hline                   
 & $V_p$, km~s$^{-1}$ & $T_p$, 10$^4$ K & $O^{7+}$/$O^{6+}$ & $Fe$/$O$ & $Q_{Fe}$ & $|B|$, nT \\
\hline
ICME & 450 & $<$~5 & $>$~0.6 & $>$~0.2 & $>$~12 & 3 -- 39 \\
\hline
CIR & $>$~450 & $>$~7 & $<$~0.1 & 0.02 -- 0.19 & 9 -- 10 & 7 -- 11 \\
\hline
Ambient SW & 360 & 6 & 0.1 & 0.1 & 9 -- 11 & 4 \\
\hline

\end{tabular}
\end{table}

The multi-source transients can be classified into several types: 1) complex ejecta with weak CME-CME interaction, 2) complex ejecta originated from strong CME-CME interactions, and 3) complex structures consisting of interacting CMEs and high-speed streams (HSS), including merged interaction regions (MIRs: \citealp{burlaga2003,rouillard2010}).

Successive or interacting CMEs may be sympathetic or homologous. In general, sympathetic CMEs originate almost simultaneously from different source regions with a certain physical connection \citep{moon2003}. Homologous CMEs occur successively from the same region in an interval of several hours and have a similar morphology \citep{zhang2002}. A review of different aspects associated with the interaction of successive CMEs can be found in the paper of \cite{lugaz17}.

In the next sections we describe signatures of single-source ICME and different multi-source complex transient events observed during the considered period of Cycle~24. The possible sources of these events are given in Table~2 and Table~3.

\subsection{Single-source event}
\label{S event}

Below we consider and date the ICMEs according to the RC list. We consider the ICME on 5 April 2010 (RC) as an example of a classic single-source event. The propagation of this ICME and its influence on Earth were analyzed in \cite{mostl2010,liu2011,temmer2011}. The most probable source of this event was the halo CME erupted from the Sun on 3 April 2010. This CME was associated with a flare from active region (AR) 11059. The time and speed of the ICME and parameters of its presumed source are given in Table~2.

The position of the source with respect to the Sun-Earth line is S27W00, so the CME propagated radially from the site of the associated flare \citep{mostl2010}. \cite{temmer2011} derived that the trajectory of the apex of this CME of E25~$\pm$~10 differs by 10--30 degrees from the results given in \cite{mostl2010}.

ACE data (Fig.~3) show the shock in front of the ICME and the sheath between them. We can see a smooth rotation of magnetic field components which is typical for MCs. This ICME triggered a prolonged geomagnetic storm on 5--7 April 2010 with the minimum Dst=~-72~nT \citep{mostl2010}. It also caused a gradual solar energetic particle (SEP) event, which was investigated in detail in \cite{liu2011}.

\begin{figure}[!h] 
\centerline{
\includegraphics[width=1.\textwidth,clip=]{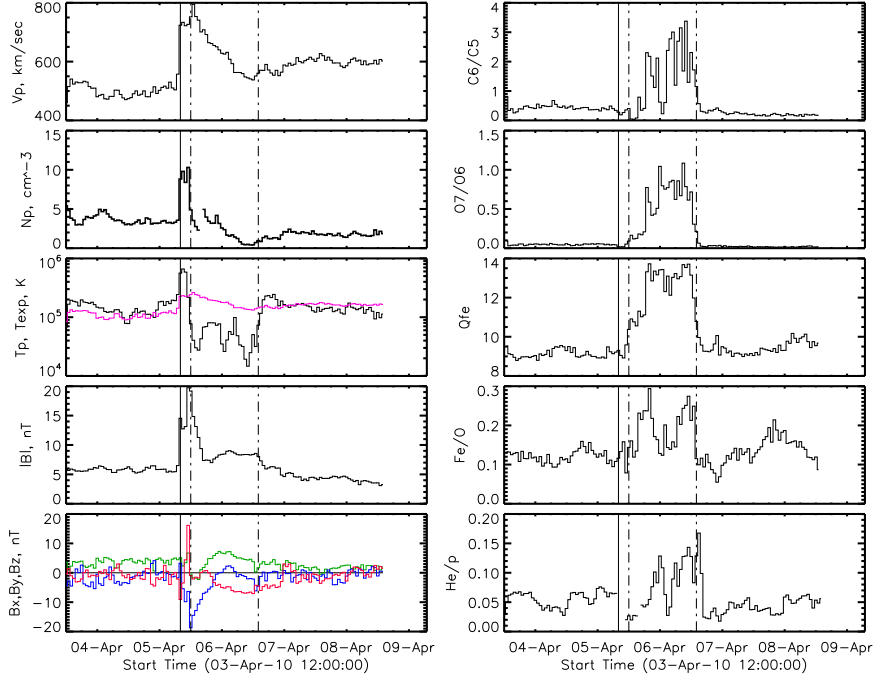}
}
\caption{Single-source event on 5 April 2010. Left panel from top to bottom: the proton speed; the proton density; the proton temperature (black) and the expected temperature (magenta); the MF intensity; the MF GSM components (Bx -- the green, By -- the blue, Bz -- the red). Right panel from top to bottom: the $C^{6+}$/$C^{5+}$ ratio; $O^{7+}$/$O^{6+}$ ratio; mean iron charge $<Q_{Fe}>$; the $Fe$/$O$ ratio; and the $He$/$p$ ratio. The dot-dashed vertical lines mark the start and end of the ICME from the RC list. Solid black vertical lines marks the shock.}
\label{fig3}
\end{figure}

The boundaries of ion charge state enhancements (Fig.~3, right panels) well agree with the boundaries of the ICME in RC list determined on the base of the main plasma parameters $V_p$, $n_p$, $T_p$ and $B$ (Fig.~3, left panels). The $Fe$/$O$ ratio has two peaks with a local depression between them (FIP bias varies between 5 and 3). As the FIP-bias depends on the magnetic field topology in the solar source region \citep{feldman1992,somov2013,laming2015}, the variation of the $Fe$/$O$ ratio suggests that the ejecta includes components of plasma with different magnetic nature. Also, the proton temperature is below the expected value which is one of the main signatures of standard single-source ICMEs (\textit{e.g.} \citealp{rich2010}).

\subsection{Multiple-source events}
\label{M event 1}

\subsubsection{Complex ejecta with weak interaction}
\label{M CMEs}

The ICME on 11 February 2010 (RC) is an example of merged successive CMEs forming complex ejecta with weak interaction. The most probable sources were three CMEs erupted from Sun on 6 -- 7 February 2010. These CMEs were associated with three flares from AR~11045. The time and speed of the ICME and parameters of its sources are given in Table~2.

\begin{figure}[!h] 
\centerline{
\includegraphics[width=1.\textwidth,clip=]{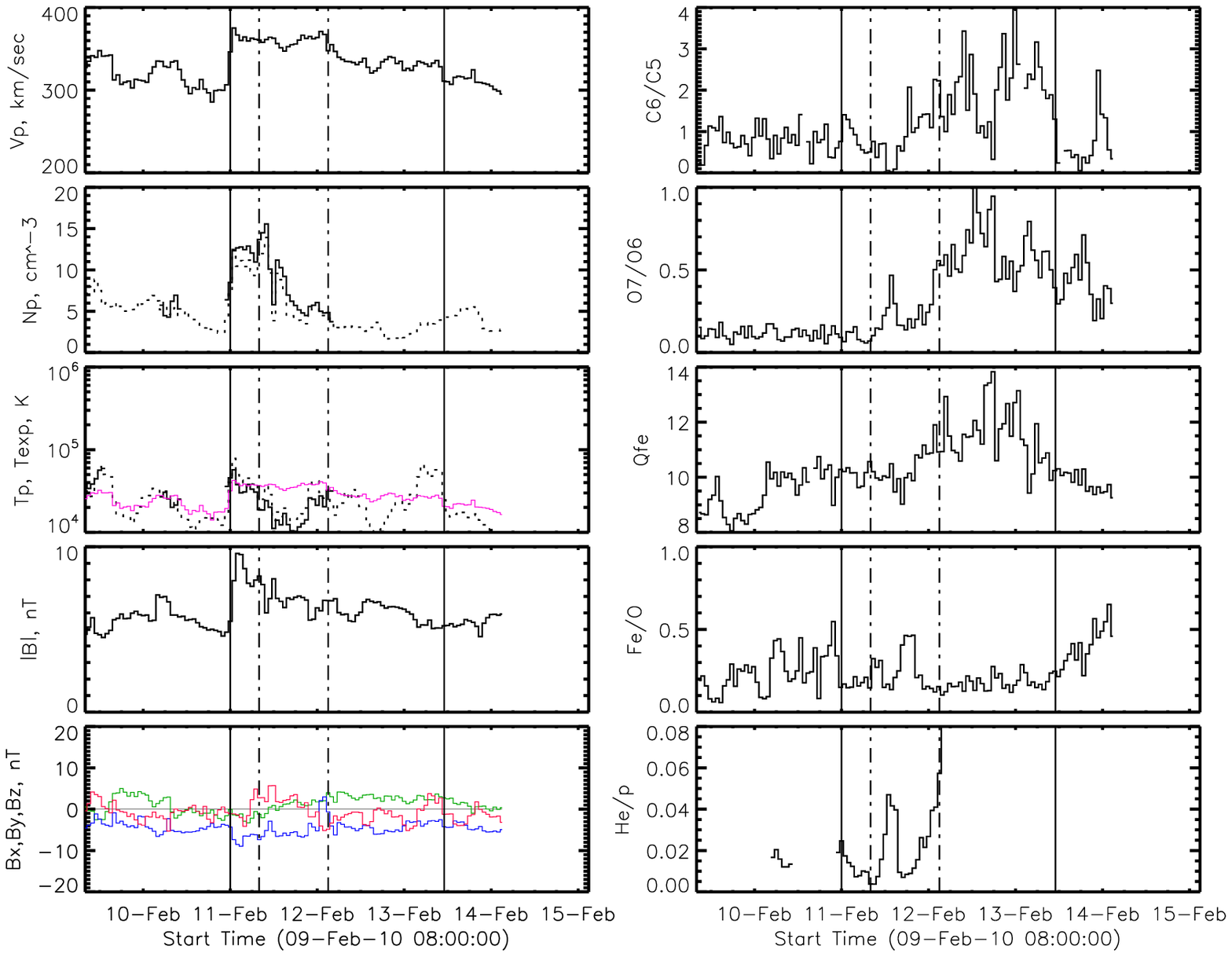}
}
\caption{Complex ejecta with weak interaction on 11 February 2010. Left panel from top to bottom: the proton speed; the proton density; the proton temperature (black) and the expected temperature (magenta); the MF module; the MF GSM components (Bx -- the green, By -- the blue, Bz -- the red). Right panel from top to bottom: the $C^{6+}$/$C^{5+}$ ratio; $O^{7+}$/$O^{6+}$ ratio; mean iron charge $<Q_{Fe}>$; the $Fe$/$O$ ratio; and the $He$/$p$ ratio. The dot-dashed vertical lines mark the start and end of ICME from the RC list. The first solid black vertical line marks the shock, the second solid vertical line mark the end of the transient event according to the ion composition data. The dotted lines show the data taken by the WIND spacecraft.}
\label{fig4}
\end{figure}

According to the RC list, the ICME lasted from 11 February 2010 8:00 UT to 12 February 2010 3:00 UT, but after the end of the ICME one can see the period of depressed proton temperature, probably associated with the second ejecta. The ion composition parameters show several maxima, the first one coinciding with the ICME boundaries from the RC list, and the other two being beyond them (Fig.~4). In these peaks the enhanced values of $O^{7+}$/$O^{6+}$ ratio (around 0.8) and $Q_{Fe}$ ($>$~12) indicate the presence of hot plasma in the solar source which is typical for ICMEs. Taking into account the ion composition distribution, the end boundary of this event can be extended to 13 February 2010 11:00 UT. Thus, we see the ion composition enhancements from successive CMEs are merged and create a long-lasting (2.5 day) complex ejecta structure whereas the ICME identified in the RC list lasted only 19 hours. We also can see the double-peak structure in $He$/$p$ ratio as support of this hypothesis. According to \citet{lugaz17}, this case of CME-CME interaction can be classified as a long-duration complex ejecta, consisting in our case of MC and ejecta. We can see only one shock probably related to the second (fastest) CME  observed close to the Sun (see Table 2). We note though that in this case it is impossible to determine reliably the correspondence between the solar CMEs and transient interplanetary structures, as it is the case in complex ejecta \citep{burlaga2002}.

\subsubsection{Complex ejecta with strong interaction}
\label{CME-CME}

The event on 6 August 2011 is an example of a complex ejecta with strong CME-CME interaction. The most possible sources of this event were two CMEs erupted from the Sun on 3 -- 4 August 2011 (see Table~3). A case study of this complex event consisting of several solar wind transients detected by ACE on 4 -- 7 August 2011 was presented by \cite{rodkin2017}. Contrary to the case of merged CMEs shown in Figure~4, one can see here one shock and one ejecta coinciding with the ICME in the RC list.

\begin{figure}[!h] 
\centerline{
\includegraphics[width=1.\textwidth,clip=]{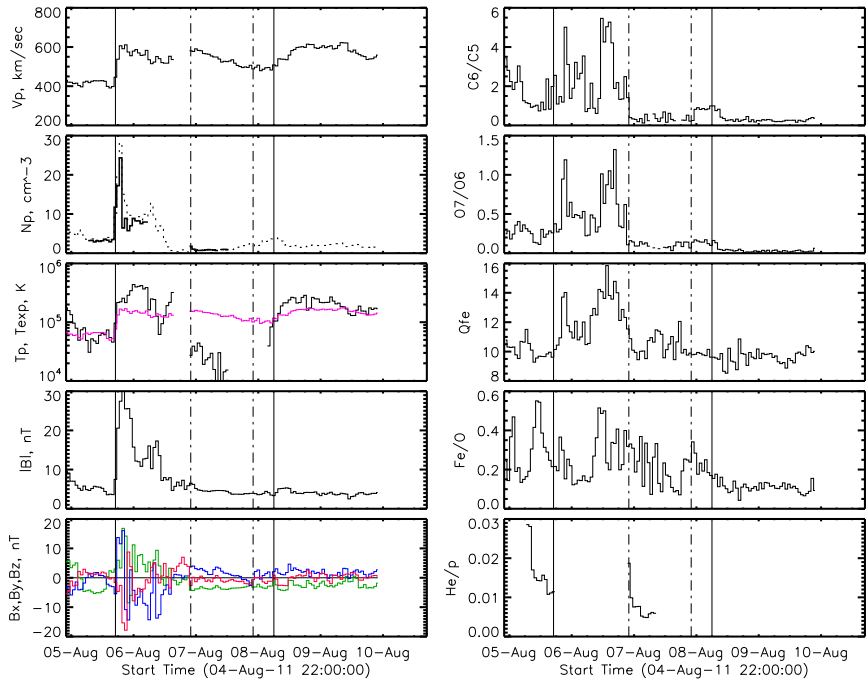}
}
\caption{Complex ejecta formed by strong CME-CME interaction on 6 August 2011. The layout and the format of the plots are the same as in Figure~4.}
\label{fig5}
\end{figure}

Figure~5 presents the record of the complex ejecta event created by strong interaction of two CMEs. The second one overtook the first at 0.6~AU \footnote{http://helioweather.net/}. As a result, it created a complex structure consisted of shock, sheath and ICME lasting from 5 to 8 August 2011. There are two ion charge state enhancements and a two-peak density enhancement. The first peak corresponds to the plasma of the first CME on 3 August 2011 compressed by the second CME on 4 August 2011, the second ion charge state enhancement most probably can be related to the second hotter CME. This event can be referred to the type of the complex ejecta \citep{lugaz17} with strong interaction between participating CMEs. The strong interaction is evidenced by the presence of ICME material (with ion charge state composition signatures) from the first CME in the sheath between the the second ICME and its driven shock.

\subsubsection{CME-HSS interaction}
\label{CME-HSS}

The event on 4 February 2011 is an example of the complex structure resulting from the CME-HSS interaction (Fig.~6). The event started with a shock detected by WIND on 4 February 2011 at 01:51~UT followed by a sheath and the ICME (the RC list: started on 4 February at 13:00~UT, ended at 20:00~UT). The most possible source of this ICME was the slow CME erupted from Sun on 30 January 2011 and observed by STEREO-A/ STEREO-B at 03:54/09:24~UT, and by LASCO at 12:36~UT (see Table~4). The CME was rather slow and weak, so due to different sensitivities of the coronagraphs, the start times differ by more than 1~hour; association of all data with the same feature was checked visually. Using the drag-based model and taking the ambient wind speed as 350~km~s$^{-1}$, the predicted time of the CME arrival to Earth is on 5 February at 12:46~UT. It means that the ICME arrived one day earlier than predicted. This difference may be caused by interaction of the CME with the HSS originated from the northern mid-latitude coronal hole, which appeared in the central part of the solar disk after 29 January 2011, achieving its maximal area on 1 February around 17:00~UT. Using the empirical model based on the coronal hole area \citep{shugay2011}, it can be calculated, that in absence of interaction the stream interface of the HSS (a boundary of the corotating interaction region -- CIR) should arrive to Earth on 3 February at 20:00~UT with the peak speed of $\approx$~600~km~s$^{-1}$.

\begin{figure}[!h] 
\centerline{
\includegraphics[width=1.\textwidth,clip=]{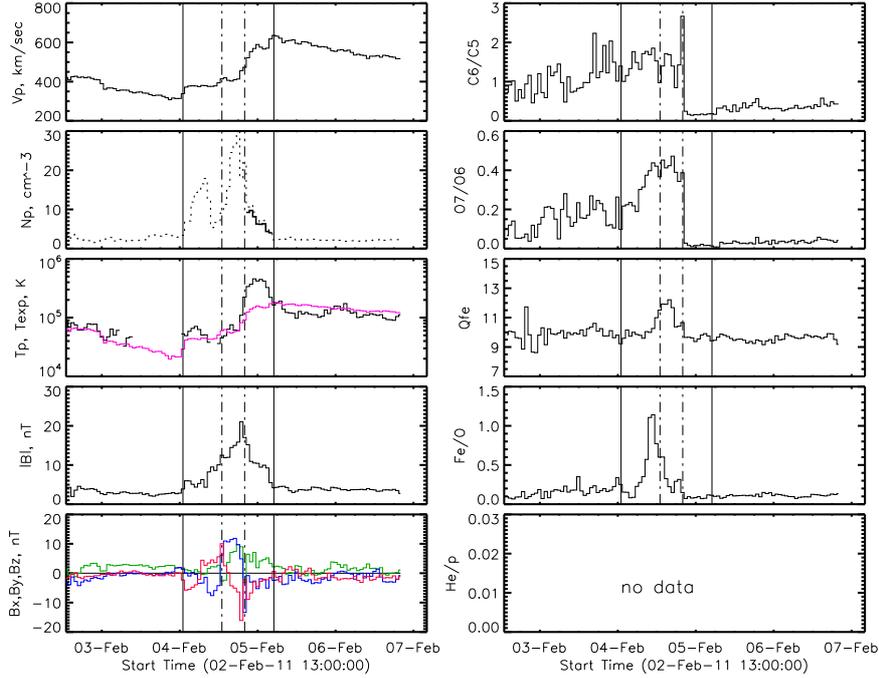}
}
\caption{MIR created by the CME-HSS interaction on 4 February 2011. The layout and the format of the plots are the same as in Figure~4.}
\label{fig6}
\end{figure}

According to the ACE data, the CIR arrived on 4 February at 20:00~UT, just after the ICME but one day later than predicted. In the ICME region, the magnetic field configuration was typical for a MC. At the moment of peak $|B|$, the $B_z$ component dropped to -16~nT, which caused a moderate geomagnetic storm at 22:00~UT of the same day with Dst~=~-63~nT. The ion composition distributions show enhancements, which started after the shock and lasted to the end of ICME. After the ICME, in the CIR region the composition signatures drop down to the values typical for the HSS plasma. It is worth to note that the peak in $Fe$/$O$ associated with the flux rope was shifted by around 6 hours earlier with respect to the peak of $<Q_{Fe}>$ and plasma density. This suggests that plasma of the leading flux rope was relatively cold in comparison with that in the following hotter CME body.

We suggest, that the observed structure of the complex event and divergence between the real and predicted arrival times of the CME and the HSS can be explained by their interaction. The CME was additionally accelerated by the following HSS from $\sim$~100~km~s$^{-1}$ in the low corona to 400~km~s$^{-1}$ at 1~AU. Due to this interaction the HSS got slowed down and its signatures appeared at 1~AU a day later than predicted. The observed complex structure can be classified as a MIR consisting of a shock, sheath, MC and a CIR similar to the case described by \citet{rouillard2010}.

\section{Summary and Conclusion}
\label{S-Conclusions}

We analyzed statistics, solar sources and properties of ICMEs and complex transient structures in the solar wind observed \textit{in situ} in the beginning of Cycle~24 from January 2010 to August 2011. The total number of ICMEs and X-ray flares during the first eight years of Cycle~24 was two times less than during the same period of Cycle~23, but the total number of CMEs was similar. Correlation of the ICME yearly frequencies with those of CMEs during Cycle~24 was similar to Cycle~23 for the considered period, whereas the correlation of ICMEs with flares decreased during Cycle~24 due to noticeable amount of weak flares below the measurable level. Consequently, the ICME statistics with respect to their solar sources has been changed due to the increased portion of fairly identified solar events and solar wind ICME-like transients \citep{kilpua2014}. A study of the ion composition of ICMEs for the period of growing activity of Cycle~24 (January 2010 -- August 2011) shows that the averaged values of $<Q_{Fe}>$ and $O^{7+}$/$O^{6+}$ ratio for ICMEs during the given period decreased in comparison with the mean values for Cycle~23 (the period 1998 -- 2000), which supports this conclusion \citep{gopals13,galvin2013,rodkin2016a}.

Another feature of Cycle~24 is a noticeable number of complex transient structures arising because of interaction between several streams in the heliosphere. We identified solar origins of ICMEs from the RC list for the period of January 2010 -- August 2011 and found that in many cases the event can be associated with two or more solar sources or with presumed interaction of CMEs with HSSs from coronal holes. We considered in detail four examples of single- and multi-source events paying particular attention to the ion composition parameters.

The cases of the SS and MS events considered in the previous sections and presented in Figures~3--6 demonstrate the relation between ion composition parameters and other properties of the transient streams, as well as their interaction in the heliosphere. Analysis of the ACE solar wind data has shown that the SS events satisfy the identification criteria established in the literature for ICMEs \citep{lepri2001,rich04,zurbuchen06,rich2010}. According to Tables~5 and 6, and Figure~2, temporal boundaries for the SS events determined by the standard identification procedure and by ion composition are similar (1.12 and 1 day, correspondingly), which agree well with the averaged ICME duration (1.27~$\pm$~0.67 days, \citealp{temmer2017}), obtained from the RC list without division on SS and MS events. For the MS events, the average total duration, determined from the ion charge composition, is 2.4 days, which is more than 2 times longer than the average SS event duration.

For different types of the ICME events, temporal profiles of the ion composition parameters are strongly different. In the SS events, the profiles of $O^{7+}$/$O^{6+}$ and $<Q_{Fe}>$ have one broad peak with the full width similar to the event duration, sometimes with local depressions with characteristic times of 4 -- 6 hours (see Fig.~3). The average values of the ion charge state parameters correspond to the plasma temperature in the solar source below the boundary of the freezing-in region. A fine structure inside these maxima is most likely linked with spatial or temporal variation in the magnetic field topology of the solar source.

In the MS complex ejecta events of the weakly interacting type (Fig.~4) the ion composition profiles can have several maxima following one after another with the width of $\sim$~1 day as for the SS events. As in this case there is very little CME plasma mixing, we may expect that the peak values of $O^{7+}$/$O^{6+}$ and $<Q_{Fe}>$ correspond to the plasma temperatures in the individual solar sources of these CMEs like it happens in the SS events. Other parameters of the transients such as speed, density and proton temperature profiles reflect the successive passing of the CMEs through the interplanetary medium. \citet{temmer2017} have shown that typically the disturbed state of the ambient solar wind relaxes during 3 -- 5 days after the ICME passage, so the propagation of subsequent CMEs following with in this time period depends on the previous ones, and the result cannot be described as a simple chain of several independent ICMEs.

The case of MS complex ejecta event associated with strong CME-CME interaction is the most complicated for interpretation. Collision of CMEs results in variation of their kinematical parameters, compression and mixing of their plasma. Kinematical and magnetic field aspects of this interaction were considered, in particular, by \citealp{mostl12,temmer12,liu2014,lugaz17,shen2017} and in the references therein. The complex MS event of 6 August 2011, presented in Figure~5, occurred after interaction of the faster and more powerful CME ($V_{CME}$=1315~km~s$^{-1}$) , which catches up with the previous slower one ($V_{CME}$=610~km~s$^{-1}$) started 14 hours earlier. Simulation by the WSA-Enlil cone model\footnote{http://www.helioweather.net} shows that this interaction occurred in the heliosphere at the distance of $\sim$~0.6 AU from the Sun. The first small peak in the $O^{7+}$/$O^{6+}$ ratio around 1.2 and in the iron ion charge state $<Q_{Fe}>$ around 14 occurred in the sheath after the shock that arrived on 5 August at 17:51 (according to the RC list), and, probably, corresponds to the plasma of the first CME compressed by the second one driving the shock. The second wider peak in composition parameters ($O^{7+}$/$O^{6+}$~$\sim$~1.3, $<Q_{Fe}>$~$\sim$~16, $Fe$/$O$~$\sim$~0.5) seen on 6 August at $\sim$~12:00~UT hours likely refers to the hottest plasma of MC originated from the second faster CME. On 6 August at 00:00~UT the GSM $B_z$ component of the IMF dropped down to -14~nT, which led to the geomagnetic storm with Dst=-110~nT at 12:00~UT. The ICME identified by the RC list between 6 August, 22:00~UT and 7 August, 22:00~UT corresponding to the interval of low proton temperature contained the plasma ($O^{7+}$/$O^{6+}$~$\sim$~0.1 -- 0.2, $<Q_{Fe}>$~$\sim$~10 -- 12), probably, created by the two merged ejecta. As plasma in the heliosphere is collisionless, its ion charge state formed in the low corona cannot be changed by the stream interaction. Thus, by analysis of the ion composition of different plasma components, we can trace the plasma streams from their solar sources.

In the cases of CME-HSS interaction the resulting complex structure (MIR) consists of several parts with the ion charge state depending of the history of this event (\textit{e.g.} the faster CME overtakes the slower HSS or vice versa). In the case presented in Figure~6, the faster HSS overtakes a CME. Due to this interaction, the ion composition profiles have broad peaks corresponding to the CME appeared earlier than the plasma signatures of ICME. The CME plasma originated from closed magnetic structure of active region has enhanced $Fe$/$O$~$\sim$~1 (FIP bias $\sim$~20), whereas the plasma of HSS from coronal hole has $Fe$/$O$~$\sim$~0.05 -- 0.1 (FIP bias $\sim$~1 -- 2). The regions of the CME and HSS plasmas are easily distinguished in the ion composition profiles.

Summarizing the features of the considered cases, we can formulate the specific properties of the ion composition in the SS and MS events:
\begin{enumerate}
\item   In the SS events, the charge state enhancements of the $O$ and $Fe$ ions in the solar wind at 1~AU coincide in time with the plasma and magnetic field signatures used for the ICME identification. The averaged ion charge states in the ICME correspond to the freezed-in conditions (plasma temperature and density) in the solar source.
\item	In the MS complex ejecta events with weak interaction of several CMEs, the total profile of the ion charge state consists of a number of enhancements associated with the successive CMEs. In absence of a strong interaction between the transients (\textit{e.g.} in case of a not too important speed difference between parent solar CMEs), the magnitudes of the composition enhancements are determined by plasma conditions in the parent solar sources, but their arrival times may be shifted due to incomplete recovery of the background interplanetary medium after the passage of the preceding transient.
\item	The ion composition profiles in the MS complex ejecta events associated with the strong CME-CME interaction at 1~AU have a complicated structure depending on many factors, as timelines and parameters of the participating CMEs, mutual orientation of their magnetic fields, presence of one or several shocks and sheaths. However, the ion charge states of the plasma components of this complex structure are defined by parameters of the solar sources and are not strongly disturbed by the interaction.
\item	In the MS events with interaction between CME and HSS, the resulting profiles of the ion composition parameters in the complex structures (MIRs) depend on the order and parameters of solar sources of the arriving streams.
\end{enumerate}

\begin{acks}
The authors are grateful to Ian Richardson and Hilary Cane for their list of Near-Earth Interplanetary Coronal Mass Ejections\footnote{http://www.srl.caltech.edu/ACE/ASC/DATA/level3/icmetable2.htm}, which we used in our investigations. Also this paper uses data from the CACTus CME catalog\footnote{http://sidc.oma.be/cactus/}, generated and maintained by the SIDC at the Royal Observatory of Belgium, and the SEEDS CME catalog\footnote{http://spaceweather.gmu.edu/seeds/}. The SEEDS project has been supported by NASA Living With a Star Program and NASA Applied Information Systems Research Program. We have used the CME catalog that is generated and maintained at the CDAW Data Center\footnote{https://cdaw.gsfc.nasa.gov/} by NASA and The Catholic University of America in cooperation with the Naval Research Laboratory. SOHO is a project of international cooperation between ESA and NASA. The authors thank the STEREO, GOES, SDO/AIA, and ACE research teams for their open data policy. We are grateful for the opportunity to use the results of the simulation obtained by the WSA-Enlil Cone and DBM models\footnote{http://helioweather.net}. This work was supported by the Russian Scientific Foundation project 17-12-01567. A. N. Zhukov acknowledges support from the Belgian Federal Science Policy Office through the ESA-PRODEX programme.
\end{acks}

\bibliographystyle{spr-mp-sola}
\bibliography{bibliography}

\begin{thebibliography}{85}
\ifx\bisbn     \undefined \def\bisbn  #1{ISBN #1}\fi
\ifx\binits    \undefined \def\binits#1{#1}\fi
\ifx\bauthor   \undefined \def\bauthor#1{#1}\fi
\ifx\batitle   \undefined \def\batitle#1{#1}\fi
\ifx\bjtitle   \undefined \def\bjtitle#1{\textit{#1}}\fi
\ifx\bvolume   \undefined \def\bvolume#1{\textbf{#1}}\fi
\ifx\byear     \undefined \def\byear#1{#1}\fi
\ifx\bissue    \undefined \def\bissue#1{#1}\fi
\ifx\bfpage    \undefined \def\bfpage#1{#1}\fi
\ifx\blpage    \undefined \def\blpage #1{#1}\fi
\ifx\burl      \undefined \def\burl#1{\textsf{#1}}\fi
\ifx\href      \undefined \def\href#1#2{\textsf{#2}}\fi
\ifx\betal     \undefined \def\betal{\textit{et al.}}\fi
\ifx\bctitle   \undefined \def\bctitle#1{#1}\fi
\ifx\beditor   \undefined \def\beditor#1{#1}\fi
\ifx\bbtitle   \undefined \def\bbtitle#1{\textit{#1}}\fi
\ifx\bedition  \undefined \def\bedition#1{#1}\fi
\ifx\bseriesno \undefined \def\bseriesno#1{\textbf{#1}}\fi
\ifx\blocation \undefined \def\blocation#1{#1}\fi
\ifx\bsertitle \undefined \def\bsertitle#1{\textit{#1}}\fi
\ifx\bsnm      \undefined \def\bsnm#1{#1}\fi
\ifx\bsuffix   \undefined \def\bsuffix#1{#1}\fi
\ifx\bparticle \undefined \def\bparticle#1{#1}\fi
\ifx\barticle  \undefined \def\barticle#1{}\fi
\ifx\binstitute  \undefined \def\binstitute#1{#1}\fi
\ifx\bpublisher  \undefined \def\bpublisher#1{#1}\fi
\ifx\doiurl    \undefined
  \def\doiurl#1{\href{http://dx.doi.org/#1}{\textsf{DOI}}}\fi
\ifx\arxivurl  \undefined
  \def\arxivurl#1{\href{http://arxiv.org/abs/#1}{\textsf{arXiv}}}\fi
\ifx\adsurl    \undefined
  \def\adsurl#1{\href{http://adsabs.harvard.edu/abs/#1}{\textsf{ADS}}}\fi
\ifx\botherref \undefined \def\botherref#1{}\fi
\ifx\url       \undefined \def\url#1{\textsf{#1}}\fi
\ifx\bchapter  \undefined \def\bchapter#1{}\fi
\ifx\bbook     \undefined \def\bbook#1{}\fi
\ifx\bcomment  \undefined \def\bcomment#1{#1}\fi
\ifx\oauthor   \undefined \def\oauthor#1{#1}\fi
\ifx\citeauthoryear \undefined\def \citeauthoryear#1{#1}\fi
\ifx\endbibitem\undefined \def\endbibitem{}\fi
\ifx\bconflocation  \undefined \def\bconflocation#1{#1} \fi

\bibitem[\protect\citeauthoryear{{Andrews}}{2003}]{andrews03}
\begin{barticle}
\bauthor{\bsnm{{Andrews}}, \binits{M.-D.-}}:
\byear{2003},
\batitle{{A Search for CMEs Associated with Big Flares}}.
\bjtitle{Solar Physics}
\bvolume{218},
\bfpage{261}.
\doiurl{10-1023/B:SOLA-0000013039-69550-bf}.
\adsurl{http://cdsads-u-strasbg-fr/abs/2003SoPh--218--261A}.
\end{barticle}
\endbibitem

\bibitem[\protect\citeauthoryear{{Behannon}, {Burlaga}, and
  {Hewish}}{1991}]{beh1991}
\begin{barticle}
\bauthor{\bsnm{{Behannon}}, \binits{K.W.}},
\bauthor{\bsnm{{Burlaga}}, \binits{L.F.}},
\bauthor{\bsnm{{Hewish}}, \binits{A.}}:
\byear{1991},
\batitle{{Structure and evolution of compound streams at not greater than 1
  AU}}.
\bjtitle{\jgr}
\bvolume{96},
\bfpage{21}.
\doiurl{10.1029/91JA02267}.
\adsurl{1991JGR....9621213B}.
\end{barticle}
\endbibitem

\bibitem[\protect\citeauthoryear{{Borovsky} and {Denton}}{2006}]{borovsky2006}
\begin{barticle}
\bauthor{\bsnm{{Borovsky}}, \binits{J.E.}},
\bauthor{\bsnm{{Denton}}, \binits{M.H.}}:
\byear{2006},
\batitle{{Differences between CME-driven storms and CIR-driven storms}}.
\bjtitle{Journal of Geophysical Research (Space Physics)}
\bvolume{111},
\bfpage{A07S08}.
\doiurl{10.1029/2005JA011447}.
\adsurl{2006JGRA..111.7S08B}.
\end{barticle}
\endbibitem

\bibitem[\protect\citeauthoryear{{Brueckner}
  \textit{et~al.}}{1995}]{brueckner95}
\begin{barticle}
\bauthor{\bsnm{{Brueckner}}, \binits{G.E.}},
\bauthor{\bsnm{{Howard}}, \binits{R.A.}},
\bauthor{\bsnm{{Koomen}}, \binits{M.J.}},
\bauthor{\bsnm{{Korendyke}}, \binits{C.M.}},
\bauthor{\bsnm{{Michels}}, \binits{D.J.}},
\bauthor{\bsnm{{Moses}}, \binits{J.D.}},
\bauthor{\bsnm{{Socker}}, \binits{D.G.}},
\bauthor{\bsnm{{Dere}}, \binits{K.P.}},
\bauthor{\bsnm{{Lamy}}, \binits{P.L.}},
\bauthor{\bsnm{{Llebaria}}, \binits{A.}},
\bauthor{\bsnm{{Bout}}, \binits{M.V.}},
\bauthor{\bsnm{{Schwenn}}, \binits{R.}},
\bauthor{\bsnm{{Simnett}}, \binits{G.M.}},
\bauthor{\bsnm{{Bedford}}, \binits{D.K.}},
\bauthor{\bsnm{{Eyles}}, \binits{C.J.}}:
\byear{1995},
\batitle{{The Large Angle Spectroscopic Coronagraph (LASCO)}}.
\bjtitle{\solphys}
\bvolume{162},
\bfpage{357}.
\doiurl{10.1007/BF00733434}.
\adsurl{http://cdsads.u-strasbg.fr/abs/1995SoPh..162..357B}.
\end{barticle}
\endbibitem

\bibitem[\protect\citeauthoryear{{Burlaga}, {Behannon}, and
  {Klein}}{1987}]{burlaga1987}
\begin{barticle}
\bauthor{\bsnm{{Burlaga}}, \binits{L.F.}},
\bauthor{\bsnm{{Behannon}}, \binits{K.W.}},
\bauthor{\bsnm{{Klein}}, \binits{L.W.}}:
\byear{1987},
\batitle{{Compound streams, magnetic clouds, and major geomagnetic storms}}.
\bjtitle{\jgr}
\bvolume{92},
\bfpage{5725}.
\doiurl{10.1029/JA092iA06p05725}.
\adsurl{1987JGR....92.5725B}.
\end{barticle}
\endbibitem

\bibitem[\protect\citeauthoryear{{Burlaga}, {Plunkett}, and
  {St.~Cyr}}{2002}]{burlaga2002}
\begin{barticle}
\bauthor{\bsnm{{Burlaga}}, \binits{L.F.}},
\bauthor{\bsnm{{Plunkett}}, \binits{S.P.}},
\bauthor{\bsnm{{St.~Cyr}}, \binits{O.C.}}:
\byear{2002},
\batitle{{Successive CMEs and complex ejecta}}.
\bjtitle{Journal of Geophysical Research (Space Physics)}
\bvolume{107},
\bfpage{1266}.
\doiurl{10.1029/2001JA000255}.
\adsurl{2002JGRA..107.1266B}.
\end{barticle}
\endbibitem

\bibitem[\protect\citeauthoryear{{Burlaga} \textit{et~al.}}{1981}]{burlaga1981}
\begin{barticle}
\bauthor{\bsnm{{Burlaga}}, \binits{L.}},
\bauthor{\bsnm{{Sittler}}, \binits{E.}},
\bauthor{\bsnm{{Mariani}}, \binits{F.}},
\bauthor{\bsnm{{Schwenn}}, \binits{R.}}:
\byear{1981},
\batitle{{Magnetic loop behind an interplanetary shock - Voyager, Helios, and
  IMP 8 observations}}.
\bjtitle{\jgr}
\bvolume{86},
\bfpage{6673}.
\doiurl{10.1029/JA086iA08p06673}.
\adsurl{1981JGR....86.6673B}.
\end{barticle}
\endbibitem

\bibitem[\protect\citeauthoryear{{Burlaga} \textit{et~al.}}{2003}]{burlaga2003}
\begin{barticle}
\bauthor{\bsnm{{Burlaga}}, \binits{L.}},
\bauthor{\bsnm{{Berdichevsky}}, \binits{D.}},
\bauthor{\bsnm{{Gopalswamy}}, \binits{N.}},
\bauthor{\bsnm{{Lepping}}, \binits{R.}},
\bauthor{\bsnm{{Zurbuchen}}, \binits{T.}}:
\byear{2003},
\batitle{{Merged interaction regions at 1 AU}}.
\bjtitle{Journal of Geophysical Research (Space Physics)}
\bvolume{108},
\bfpage{1425}.
\doiurl{10.1029/2003JA010088}.
\adsurl{2003JGRA..108.1425B}.
\end{barticle}
\endbibitem

\bibitem[\protect\citeauthoryear{{Cane} and {Richardson}}{2003}]{cane2003}
\begin{barticle}
\bauthor{\bsnm{{Cane}}, \binits{H.V.}},
\bauthor{\bsnm{{Richardson}}, \binits{I.G.}}:
\byear{2003},
\batitle{{Interplanetary coronal mass ejections in the near-Earth solar wind
  during 1996-2002}}.
\bjtitle{Journal of Geophysical Research (Space Physics)}
\bvolume{108},
\bfpage{1156}.
\doiurl{10.1029/2002JA009817}.
\adsurl{2003JGRA..108.1156C}.
\end{barticle}
\endbibitem

\bibitem[\protect\citeauthoryear{{Compagnino}, {Romano}, and
  {Zuccarello}}{2016}]{comp16}
\begin{botherref}
\oauthor{\bsnm{{Compagnino}}, \binits{A.}},
\oauthor{\bsnm{{Romano}}, \binits{P.}},
\oauthor{\bsnm{{Zuccarello}}, \binits{F.}}:
2016,
{A statistical study of CME properties and of the correlation between flares
  and CMEs over the solar cycles 23 and 24}.
\textit{ArXiv e-prints}.
\adsurl{2016arXiv160908943C}.
\end{botherref}
\endbibitem

\bibitem[\protect\citeauthoryear{{Delaboudini{\`e}re}
  \textit{et~al.}}{1995}]{delab1995}
\begin{barticle}
\bauthor{\bsnm{{Delaboudini{\`e}re}}, \binits{J.-P.}},
\bauthor{\bsnm{{Artzner}}, \binits{G.E.}},
\bauthor{\bsnm{{Brunaud}}, \binits{J.}},
\bauthor{\bsnm{{Gabriel}}, \binits{A.H.}},
\bauthor{\bsnm{{Hochedez}}, \binits{J.F.}},
\bauthor{\bsnm{{Millier}}, \binits{F.}},
\bauthor{\bsnm{{Song}}, \binits{X.Y.}},
\bauthor{\bsnm{{Au}}, \binits{B.}},
\bauthor{\bsnm{{Dere}}, \binits{K.P.}},
\bauthor{\bsnm{{Howard}}, \binits{R.A.}},
\bauthor{\bsnm{{Kreplin}}, \binits{R.}},
\bauthor{\bsnm{{Michels}}, \binits{D.J.}},
\bauthor{\bsnm{{Moses}}, \binits{J.D.}},
\bauthor{\bsnm{{Defise}}, \binits{J.M.}},
\bauthor{\bsnm{{Jamar}}, \binits{C.}},
\bauthor{\bsnm{{Rochus}}, \binits{P.}},
\bauthor{\bsnm{{Chauvineau}}, \binits{J.P.}},
\bauthor{\bsnm{{Marioge}}, \binits{J.P.}},
\bauthor{\bsnm{{Catura}}, \binits{R.C.}},
\bauthor{\bsnm{{Lemen}}, \binits{J.R.}},
\bauthor{\bsnm{{Shing}}, \binits{L.}},
\bauthor{\bsnm{{Stern}}, \binits{R.A.}},
\bauthor{\bsnm{{Gurman}}, \binits{J.B.}},
\bauthor{\bsnm{{Neupert}}, \binits{W.M.}},
\bauthor{\bsnm{{Maucherat}}, \binits{A.}},
\bauthor{\bsnm{{Clette}}, \binits{F.}},
\bauthor{\bsnm{{Cugnon}}, \binits{P.}},
\bauthor{\bsnm{{van Dessel}}, \binits{E.L.}}:
\byear{1995},
\batitle{{EIT: Extreme-Ultraviolet Imaging Telescope for the SOHO Mission}}.
\bjtitle{\solphys}
\bvolume{162},
\bfpage{291}.
\doiurl{10.1007/BF00733432}.
\adsurl{1995SoPh..162..291D}.
\end{barticle}
\endbibitem

\bibitem[\protect\citeauthoryear{{Echer}, {Alves}, and
  {Gonzalez}}{2005}]{echer2005}
\begin{barticle}
\bauthor{\bsnm{{Echer}}, \binits{E.}},
\bauthor{\bsnm{{Alves}}, \binits{M.V.}},
\bauthor{\bsnm{{Gonzalez}}, \binits{W.D.}}:
\byear{2005},
\batitle{{A statistical study of magnetic cloud parameters and
  geoeffectiveness}}.
\bjtitle{Journal of Atmospheric and Solar-Terrestrial Physics}
\bvolume{67},
\bfpage{839}.
\doiurl{10.1016/j.jastp.2005.02.010}.
\adsurl{2005JASTP..67..839E}.
\end{barticle}
\endbibitem

\bibitem[\protect\citeauthoryear{{Emslie} \textit{et~al.}}{2012}]{emslie2012}
\begin{barticle}
\bauthor{\bsnm{{Emslie}}, \binits{A.G.}},
\bauthor{\bsnm{{Dennis}}, \binits{B.R.}},
\bauthor{\bsnm{{Shih}}, \binits{A.Y.}},
\bauthor{\bsnm{{Chamberlin}}, \binits{P.C.}},
\bauthor{\bsnm{{Mewaldt}}, \binits{R.A.}},
\bauthor{\bsnm{{Moore}}, \binits{C.S.}},
\bauthor{\bsnm{{Share}}, \binits{G.H.}},
\bauthor{\bsnm{{Vourlidas}}, \binits{A.}},
\bauthor{\bsnm{{Welsch}}, \binits{B.T.}}:
\byear{2012},
\batitle{{Global Energetics of Thirty-eight Large Solar Eruptive Events}}.
\bjtitle{\apj}
\bvolume{759},
\bfpage{71}.
\doiurl{10.1088/0004-637X/759/1/71}.
\adsurl{2012ApJ...759...71E}.
\end{barticle}
\endbibitem

\bibitem[\protect\citeauthoryear{{Farrugia}, {Burlaga}, and
  {Lepping}}{1997}]{farrugia97}
\begin{bchapter}
\bauthor{\bsnm{{Farrugia}}, \binits{C.J.}},
\bauthor{\bsnm{{Burlaga}}, \binits{L.F.}},
\bauthor{\bsnm{{Lepping}}, \binits{R.P.}}:
\byear{1997},
\bctitle{{Magnetic Clouds and the quiet-storm effect at Earth}}.
In: \beditor{\bsnm{{Tzurutani}}, \binits{B.T.}},
\beditor{\bsnm{{Gonzalez}}, \binits{W.D.}},
\beditor{\bsnm{{Kamide}}, \binits{Y.}},
\beditor{\bsnm{{Arballo}}, \binits{J.K.}} (eds.)
\bbtitle{Magnetic Storms},
\bpublisher{Geophy. Mon. Ser. 98, AGU}, \blocation{???},
\bfpage{91}.
\end{bchapter}
\endbibitem

\bibitem[\protect\citeauthoryear{{Feldman}}{1992}]{feldman1992}
\begin{barticle}
\bauthor{\bsnm{{Feldman}}, \binits{U.}}:
\byear{1992},
\batitle{{Elemental abundances in the upper solar atmosphere}}.
\bjtitle{Physica Scripta Volume T}
\bvolume{46},
\bfpage{202}.
\adsurl{1992PhST...46..202F}.
\end{barticle}
\endbibitem

\bibitem[\protect\citeauthoryear{{Feldman}, {Landi}, and
  {Schwadron}}{2005}]{feldman05}
\begin{barticle}
\bauthor{\bsnm{{Feldman}}, \binits{U.-}},
\bauthor{\bsnm{{Landi}}, \binits{E.-}},
\bauthor{\bsnm{{Schwadron}}, \binits{N.-A.-}}:
\byear{2005},
\batitle{{On the sources of fast and slow solar wind}}.
\bjtitle{Journal of Geophysical Research (Space Physics)}
\bvolume{110},
\bfpage{A07109}.
\doiurl{10-1029/2004JA010918}.
\adsurl{http://cdsads-u-strasbg-fr/abs/2005JGRA--110-7109F}.
\end{barticle}
\endbibitem

\bibitem[\protect\citeauthoryear{{Galvin} \textit{et~al.}}{2009}]{galvin2009}
\begin{barticle}
\bauthor{\bsnm{{Galvin}}, \binits{A.B.}},
\bauthor{\bsnm{{Popecki}}, \binits{M.A.}},
\bauthor{\bsnm{{Simunac}}, \binits{K.D.C.}},
\bauthor{\bsnm{{Kistler}}, \binits{L.M.}},
\bauthor{\bsnm{{Ellis}}, \binits{L.}},
\bauthor{\bsnm{{Barry}}, \binits{J.}},
\bauthor{\bsnm{{Berger}}, \binits{L.}},
\bauthor{\bsnm{{Blush}}, \binits{L.M.}},
\bauthor{\bsnm{{Bochsler}}, \binits{P.}},
\bauthor{\bsnm{{Farrugia}}, \binits{C.J.}},
\bauthor{\bsnm{{Jian}}, \binits{L.K.}},
\bauthor{\bsnm{{Kilpua}}, \binits{E.K.J.}},
\bauthor{\bsnm{{Klecker}}, \binits{B.}},
\bauthor{\bsnm{{Lee}}, \binits{M.}},
\bauthor{\bsnm{{Liu}}, \binits{Y.C.-M.}},
\bauthor{\bsnm{{Luhmann}}, \binits{J.L.}},
\bauthor{\bsnm{{Moebius}}, \binits{E.}},
\bauthor{\bsnm{{Opitz}}, \binits{A.}},
\bauthor{\bsnm{{Russell}}, \binits{C.T.}},
\bauthor{\bsnm{{Thompson}}, \binits{B.}},
\bauthor{\bsnm{{Wimmer-Schweingruber}}, \binits{R.F.}},
\bauthor{\bsnm{{Wurz}}, \binits{P.}}:
\byear{2009},
\batitle{{Solar wind ion trends and signatures: STEREO PLASTIC observations
  approaching solar minimum}}.
\bjtitle{Annales Geophysicae}
\bvolume{27},
\bfpage{3909}.
\doiurl{10.5194/angeo-27-3909-2009}.
\adsurl{2009AnGeo..27.3909G}.
\end{barticle}
\endbibitem

\bibitem[\protect\citeauthoryear{{Galvin} \textit{et~al.}}{2013}]{galvin2013}
\begin{barticle}
\bauthor{\bsnm{{Galvin}}, \binits{A.B.}},
\bauthor{\bsnm{{Simunac}}, \binits{K.D.C.}},
\bauthor{\bsnm{{Jian}}, \binits{L.K.}},
\bauthor{\bsnm{{Farrugia}}, \binits{C.J.}},
\bauthor{\bsnm{{Popecki}}, \binits{M.A.}}:
\byear{2013},
\batitle{{Solar wind ion observations: Comparison from the depths of solar
  minimum to the rising of the cycle}}.
\bjtitle{Solar Wind 13}
\bvolume{1539},
\bfpage{15}.
\doiurl{10.1063/1.4810978}.
\adsurl{2013AIPC.1539...15G}.
\end{barticle}
\endbibitem

\bibitem[\protect\citeauthoryear{{Gloeckler}
  \textit{et~al.}}{1998}]{gloecker98}
\begin{barticle}
\bauthor{\bsnm{{Gloeckler}}, \binits{G.}},
\bauthor{\bsnm{{Cain}}, \binits{J.}},
\bauthor{\bsnm{{Ipavich}}, \binits{F.M.}},
\bauthor{\bsnm{{Tums}}, \binits{E.O.}},
\bauthor{\bsnm{{Bedini}}, \binits{P.}},
\bauthor{\bsnm{{Fisk}}, \binits{L.A.}},
\bauthor{\bsnm{{Zurbuchen}}, \binits{T.H.}},
\bauthor{\bsnm{{Bochsler}}, \binits{P.}},
\bauthor{\bsnm{{Fischer}}, \binits{J.}},
\bauthor{\bsnm{{Wimmer-Schweingruber}}, \binits{R.F.}},
\bauthor{\bsnm{{Geiss}}, \binits{J.}},
\bauthor{\bsnm{{Kallenbach}}, \binits{R.}}:
\byear{1998},
\batitle{{Investigation of the composition of solar and interstellar matter
  using solar wind and pickup ion measurements with SWICS and SWIMS on the ACE
  spacecraft}}.
\bjtitle{\ssr}
\bvolume{86},
\bfpage{497}.
\doiurl{10.1023/A:1005036131689}.
\adsurl{1998SSRv...86..497G}.
\end{barticle}
\endbibitem

\bibitem[\protect\citeauthoryear{{Gopalswamy} \textit{et~al.}}{2013}]{gopals13}
\begin{barticle}
\bauthor{\bsnm{{Gopalswamy}}, \binits{N.}},
\bauthor{\bsnm{{M{\"a}kel{\"a}}}, \binits{P.}},
\bauthor{\bsnm{{Akiyama}}, \binits{S.}},
\bauthor{\bsnm{{Xie}}, \binits{H.}},
\bauthor{\bsnm{{Yashiro}}, \binits{S.}},
\bauthor{\bsnm{{Reinard}}, \binits{A.A.}}:
\byear{2013},
\batitle{{The Solar Connection of Enhanced Heavy Ion Charge States in the
  Interplanetary Medium: Implications for the Flux-Rope Structure of CMEs}}.
\bjtitle{Solar Physics}
\bvolume{284},
\bfpage{17}.
\doiurl{10.1007/s11207-012-0215-2}.
\adsurl{2013SoPh..284...17G}.
\end{barticle}
\endbibitem

\bibitem[\protect\citeauthoryear{{Gopalswamy} \textit{et~al.}}{2015}]{gopals15}
\begin{barticle}
\bauthor{\bsnm{{Gopalswamy}}, \binits{N.}},
\bauthor{\bsnm{{Yashiro}}, \binits{S.}},
\bauthor{\bsnm{{Xie}}, \binits{H.}},
\bauthor{\bsnm{{Akiyama}}, \binits{S.}},
\bauthor{\bsnm{{M{\"a}kel{\"a}}}, \binits{P.}}:
\byear{2015},
\batitle{{Properties and geoeffectiveness of magnetic clouds during solar
  cycles 23 and 24}}.
\bjtitle{Journal of Geophysical Research (Space Physics)}
\bvolume{120},
\bfpage{9221}.
\doiurl{10.1002/2015JA021446}.
\adsurl{2015JGRA..120.9221G}.
\end{barticle}
\endbibitem

\bibitem[\protect\citeauthoryear{{Gosling} \textit{et~al.}}{1990}]{gosling90}
\begin{barticle}
\bauthor{\bsnm{{Gosling}}, \binits{J.T.}},
\bauthor{\bsnm{{Bame}}, \binits{S.J.}},
\bauthor{\bsnm{{McComas}}, \binits{D.J.}},
\bauthor{\bsnm{{Phillips}}, \binits{J.L.}}:
\byear{1990},
\batitle{{Coronal mass ejections and large geomagnetic storms}}.
\bjtitle{Geophysical Research Letters}
\bvolume{17},
\bfpage{901}.
\doiurl{10.1029/GL017i007p00901}.
\adsurl{http://cdsads.u-strasbg.fr/abs/1990GeoRL..17..901G}.
\end{barticle}
\endbibitem

\bibitem[\protect\citeauthoryear{{Gosling} \textit{et~al.}}{1991}]{gosling91}
\begin{barticle}
\bauthor{\bsnm{{Gosling}}, \binits{J.-T.-}},
\bauthor{\bsnm{{McComas}}, \binits{D.-J.-}},
\bauthor{\bsnm{{Phillips}}, \binits{J.-L.-}},
\bauthor{\bsnm{{Bame}}, \binits{S.-J.-}}:
\byear{1991},
\batitle{{Geomagnetic activity associated with earth passage of interplanetary
  shock disturbances and coronal mass ejections}}.
\bjtitle{Journal of Geophysical Research}
\bvolume{96},
\bfpage{7831}.
\doiurl{10-1029/91JA00316}.
\adsurl{http://cdsads-u-strasbg-fr/abs/1991JGR----96-7831G}.
\end{barticle}
\endbibitem

\bibitem[\protect\citeauthoryear{{Harrison} \textit{et~al.}}{2012}]{harrison12}
\begin{barticle}
\bauthor{\bsnm{{Harrison}}, \binits{R.A.}},
\bauthor{\bsnm{{Davies}}, \binits{J.A.}},
\bauthor{\bsnm{{M{\"o}stl}}, \binits{C.}},
\bauthor{\bsnm{{Liu}}, \binits{Y.}},
\bauthor{\bsnm{{Temmer}}, \binits{M.}},
\bauthor{\bsnm{{Bisi}}, \binits{M.M.}},
\bauthor{\bsnm{{Eastwood}}, \binits{J.P.}},
\bauthor{\bsnm{{de Koning}}, \binits{C.A.}},
\bauthor{\bsnm{{Nitta}}, \binits{N.}},
\bauthor{\bsnm{{Rollett}}, \binits{T.}},
\bauthor{\bsnm{{Farrugia}}, \binits{C.J.}},
\bauthor{\bsnm{{Forsyth}}, \binits{R.J.}},
\bauthor{\bsnm{{Jackson}}, \binits{B.V.}},
\bauthor{\bsnm{{Jensen}}, \binits{E.A.}},
\bauthor{\bsnm{{Kilpua}}, \binits{E.K.J.}},
\bauthor{\bsnm{{Odstrcil}}, \binits{D.}},
\bauthor{\bsnm{{Webb}}, \binits{D.F.}}:
\byear{2012},
\batitle{{An Analysis of the Origin and Propagation of the Multiple Coronal
  Mass Ejections of 2010 August 1}}.
\bjtitle{\apj}
\bvolume{750},
\bfpage{45}.
\doiurl{10.1088/0004-637X/750/1/45}.
\adsurl{2012ApJ...750...45H}.
\end{barticle}
\endbibitem

\bibitem[\protect\citeauthoryear{{Heidrich-Meisner}
  \textit{et~al.}}{2016}]{hm2016}
\begin{barticle}
\bauthor{\bsnm{{Heidrich-Meisner}}, \binits{V.}},
\bauthor{\bsnm{{Peleikis}}, \binits{T.}},
\bauthor{\bsnm{{Kruse}}, \binits{M.}},
\bauthor{\bsnm{{Berger}}, \binits{L.}},
\bauthor{\bsnm{{Wimmer-Schweingruber}}, \binits{R.}}:
\byear{2016},
\batitle{{Observations of high and low Fe charge states in individual solar
  wind streams with coronal-hole origin}}.
\bjtitle{\aap}
\bvolume{593},
\bfpage{A70}.
\doiurl{10.1051/0004-6361/201527998}.
\adsurl{2016A\%26A...593A..70H}.
\end{barticle}
\endbibitem

\bibitem[\protect\citeauthoryear{{Hess} and {Zhang}}{2017}]{hess2017}
\begin{barticle}
\bauthor{\bsnm{{Hess}}, \binits{P.}},
\bauthor{\bsnm{{Zhang}}, \binits{J.}}:
\byear{2017},
\batitle{{A Study of the Earth-Affecting CMEs of Solar Cycle 24}}.
\bjtitle{\solphys}
\bvolume{292},
\bfpage{80}.
\doiurl{10.1007/s11207-017-1099-y}.
\adsurl{2017SoPh..292...80H}.
\end{barticle}
\endbibitem

\bibitem[\protect\citeauthoryear{{Howard} \textit{et~al.}}{2008}]{howard08}
\begin{barticle}
\bauthor{\bsnm{{Howard}}, \binits{R.A.}},
\bauthor{\bsnm{{Moses}}, \binits{J.D.}},
\bauthor{\bsnm{{Vourlidas}}, \binits{A.}},
\bauthor{\bsnm{{Newmark}}, \binits{J.S.}},
\bauthor{\bsnm{{Socker}}, \binits{D.G.}},
\bauthor{\bsnm{{Plunkett}}, \binits{S.P.}},
\bauthor{\bparticle{et} \bsnm{al}}:
\byear{2008},
\batitle{{Sun Earth Connection Coronal and Heliospheric Investigation
  (SECCHI)}}.
\bjtitle{\ssr}
\bvolume{136},
\bfpage{67}.
\doiurl{10.1007/s11214-008-9341-4}.
\adsurl{http://cdsads.u-strasbg.fr/abs/2008SSRv..136...67H}.
\end{barticle}
\endbibitem

\bibitem[\protect\citeauthoryear{{Hudson} and {Cliver}}{2001}]{hudson2001}
\begin{barticle}
\bauthor{\bsnm{{Hudson}}, \binits{H.S.}},
\bauthor{\bsnm{{Cliver}}, \binits{E.W.}}:
\byear{2001},
\batitle{{Observing coronal mass ejections without coronagraphs}}.
\bjtitle{\jgr}
\bvolume{106},
\bfpage{25199}.
\doiurl{10.1029/2000JA904026}.
\adsurl{2001JGR...10625199H}.
\end{barticle}
\endbibitem

\bibitem[\protect\citeauthoryear{{Hundhausen}, {Gilbert}, and
  {Bame}}{1968}]{hundhausen68}
\begin{barticle}
\bauthor{\bsnm{{Hundhausen}}, \binits{A.-J.-}},
\bauthor{\bsnm{{Gilbert}}, \binits{H.-E.-}},
\bauthor{\bsnm{{Bame}}, \binits{S.-J.-}}:
\byear{1968},
\batitle{{Ionization State of the Interplanetary Plasma}}.
\bjtitle{J- Geophys- Res}
\bvolume{73},
\bfpage{5485}.
\doiurl{10-1029/JA073i017p05485}.
\end{barticle}
\endbibitem

\bibitem[\protect\citeauthoryear{{Kataoka} \textit{et~al.}}{2015}]{kataoka15}
\begin{barticle}
\bauthor{\bsnm{{Kataoka}}, \binits{R.}},
\bauthor{\bsnm{{Shiota}}, \binits{D.}},
\bauthor{\bsnm{{Kilpua}}, \binits{E.}},
\bauthor{\bsnm{{Keika}}, \binits{K.}}:
\byear{2015},
\batitle{{Pileup accident hypothesis of magnetic storm on 17 March 2015}}.
\bjtitle{\grl}
\bvolume{42},
\bfpage{5155}.
\doiurl{10.1002/2015GL064816}.
\adsurl{2015GeoRL..42.5155K}.
\end{barticle}
\endbibitem

\bibitem[\protect\citeauthoryear{{Kilpua} \textit{et~al.}}{2014}]{kilpua2014}
\begin{barticle}
\bauthor{\bsnm{{Kilpua}}, \binits{E.K.J.}},
\bauthor{\bsnm{{Mierla}}, \binits{M.}},
\bauthor{\bsnm{{Zhukov}}, \binits{A.N.}},
\bauthor{\bsnm{{Rodriguez}}, \binits{L.}},
\bauthor{\bsnm{{Vourlidas}}, \binits{A.}},
\bauthor{\bsnm{{Wood}}, \binits{B.}}:
\byear{2014},
\batitle{{Solar Sources of Interplanetary Coronal Mass Ejections During the
  Solar Cycle 23/24 Minimum}}.
\bjtitle{\solphys}
\bvolume{289},
\bfpage{3773}.
\doiurl{10.1007/s11207-014-0552-4}.
\adsurl{2014SoPh..289.3773K}.
\end{barticle}
\endbibitem

\bibitem[\protect\citeauthoryear{{Kocher} \textit{et~al.}}{2017}]{kocher2017}
\begin{barticle}
\bauthor{\bsnm{{Kocher}}, \binits{M.}},
\bauthor{\bsnm{{Lepri}}, \binits{S.T.}},
\bauthor{\bsnm{{Landi}}, \binits{E.}},
\bauthor{\bsnm{{Zhao}}, \binits{L.}},
\bauthor{\bsnm{{Manchester}}, \binits{W.B.} \bsuffix{IV}}:
\byear{2017},
\batitle{{Anatomy of Depleted Interplanetary Coronal Mass Ejections}}.
\bjtitle{\apj}
\bvolume{834},
\bfpage{147}.
\doiurl{10.3847/1538-4357/834/2/147}.
\adsurl{2017ApJ...834..147K}.
\end{barticle}
\endbibitem

\bibitem[\protect\citeauthoryear{{Laming}}{2015}]{laming2015}
\begin{barticle}
\bauthor{\bsnm{{Laming}}, \binits{J.M.}}:
\byear{2015},
\batitle{{The FIP and Inverse FIP Effects in Solar and Stellar Coronae}}.
\bjtitle{Living Reviews in Solar Physics}
\bvolume{12},
\bfpage{2}.
\doiurl{10.1007/lrsp-2015-2}.
\adsurl{2015LRSP...12....2L}.
\end{barticle}
\endbibitem

\bibitem[\protect\citeauthoryear{{Lawrance} \textit{et~al.}}{2016}]{lawrance16}
\begin{barticle}
\bauthor{\bsnm{{Lawrance}}, \binits{M.B.}},
\bauthor{\bsnm{{Shanmugaraju}}, \binits{A.}},
\bauthor{\bsnm{{Moon}}, \binits{Y.-J.}},
\bauthor{\bsnm{{Ibrahim}}, \binits{M.S.}},
\bauthor{\bsnm{{Umapathy}}, \binits{S.}}:
\byear{2016},
\batitle{{Relationships Between Interplanetary Coronal Mass Ejection
  Characteristics and Geoeffectiveness in the Rising Phase of Solar Cycles 23
  and 24}}.
\bjtitle{Solar Physics}
\bvolume{291},
\bfpage{1547}.
\doiurl{10.1007/s11207-016-0911-4}.
\adsurl{2016SoPh..291.1547L}.
\end{barticle}
\endbibitem

\bibitem[\protect\citeauthoryear{{Lemen} \textit{et~al.}}{2012}]{lemen12}
\begin{barticle}
\bauthor{\bsnm{{Lemen}}, \binits{J.R.}},
\bauthor{\bsnm{{Title}}, \binits{A.M.}},
\bauthor{\bsnm{{Akin}}, \binits{D.J.}},
\bauthor{\bsnm{{Boerner}}, \binits{P.F.}},
\bauthor{\bsnm{{Chou}}, \binits{C.}},
\bauthor{\bsnm{{Drake}}, \binits{J.F.}},
\bauthor{\bparticle{et} \bsnm{al}}:
\byear{2012},
\batitle{{The Atmospheric Imaging Assembly (AIA) on the Solar Dynamics
  Observatory (SDO)}}.
\bjtitle{\solphys}
\bvolume{275},
\bfpage{17}.
\doiurl{10.1007/s11207-011-9776-8}.
\adsurl{http://cdsads.u-strasbg.fr/abs/2012SoPh..275...17L}.
\end{barticle}
\endbibitem

\bibitem[\protect\citeauthoryear{{Lepri} \textit{et~al.}}{2001}]{lepri2001}
\begin{barticle}
\bauthor{\bsnm{{Lepri}}, \binits{S.T.}},
\bauthor{\bsnm{{Zurbuchen}}, \binits{T.H.}},
\bauthor{\bsnm{{Fisk}}, \binits{L.A.}},
\bauthor{\bsnm{{Richardson}}, \binits{I.G.}},
\bauthor{\bsnm{{Cane}}, \binits{H.V.}},
\bauthor{\bsnm{{Gloeckler}}, \binits{G.}}:
\byear{2001},
\batitle{{Iron charge distribution as an identifier of interplanetary coronal
  mass ejections}}.
\bjtitle{\jgr}
\bvolume{106},
\bfpage{29231}.
\doiurl{10.1029/2001JA000014}.
\adsurl{2001JGR...10629231L}.
\end{barticle}
\endbibitem

\bibitem[\protect\citeauthoryear{{Liu} \textit{et~al.}}{2014a}]{liu14}
\begin{barticle}
\bauthor{\bsnm{{Liu}}, \binits{Y.D.}},
\bauthor{\bsnm{{Yang}}, \binits{Z.}},
\bauthor{\bsnm{{Wang}}, \binits{R.}},
\bauthor{\bsnm{{Luhmann}}, \binits{J.G.}},
\bauthor{\bsnm{{Richardson}}, \binits{J.D.}},
\bauthor{\bsnm{{Lugaz}}, \binits{N.}}:
\byear{2014}a,
\batitle{{Sun-to-Earth Characteristics of Two Coronal Mass Ejections
  Interacting Near 1 AU: Formation of a Complex Ejecta and Generation of a
  Two-step Geomagnetic Storm}}.
\bjtitle{\apjl}
\bvolume{793},
\bfpage{L41}.
\doiurl{10.1088/2041-8205/793/2/L41}.
\adsurl{2014ApJ...793L..41L}.
\end{barticle}
\endbibitem

\bibitem[\protect\citeauthoryear{{Liu} \textit{et~al.}}{2014b}]{liu2014}
\begin{barticle}
\bauthor{\bsnm{{Liu}}, \binits{Y.D.}},
\bauthor{\bsnm{{Yang}}, \binits{Z.}},
\bauthor{\bsnm{{Wang}}, \binits{R.}},
\bauthor{\bsnm{{Luhmann}}, \binits{J.G.}},
\bauthor{\bsnm{{Richardson}}, \binits{J.D.}},
\bauthor{\bsnm{{Lugaz}}, \binits{N.}}:
\byear{2014}b,
\batitle{{Sun-to-Earth Characteristics of Two Coronal Mass Ejections
  Interacting Near 1 AU: Formation of a Complex Ejecta and Generation of a
  Two-step Geomagnetic Storm}}.
\bjtitle{\apjl}
\bvolume{793},
\bfpage{L41}.
\doiurl{10.1088/2041-8205/793/2/L41}.
\adsurl{2014ApJ...793L..41L}.
\end{barticle}
\endbibitem

\bibitem[\protect\citeauthoryear{{Liu} \textit{et~al.}}{2015}]{liu15}
\begin{barticle}
\bauthor{\bsnm{{Liu}}, \binits{Y.D.}},
\bauthor{\bsnm{{Hu}}, \binits{H.}},
\bauthor{\bsnm{{Wang}}, \binits{R.}},
\bauthor{\bsnm{{Yang}}, \binits{Z.}},
\bauthor{\bsnm{{Zhu}}, \binits{B.}},
\bauthor{\bsnm{{Liu}}, \binits{Y.A.}},
\bauthor{\bsnm{{Luhmann}}, \binits{J.G.}},
\bauthor{\bsnm{{Richardson}}, \binits{J.D.}}:
\byear{2015},
\batitle{{Plasma and Magnetic Field Characteristics of Solar Coronal Mass
  Ejections in Relation to Geomagnetic Storm Intensity and Variability}}.
\bjtitle{\apjl}
\bvolume{809},
\bfpage{L34}.
\doiurl{10.1088/2041-8205/809/2/L34}.
\adsurl{2015ApJ...809L..34L}.
\end{barticle}
\endbibitem

\bibitem[\protect\citeauthoryear{{Liu} \textit{et~al.}}{2011}]{liu2011}
\begin{barticle}
\bauthor{\bsnm{{Liu}}, \binits{Y.}},
\bauthor{\bsnm{{Luhmann}}, \binits{J.G.}},
\bauthor{\bsnm{{Bale}}, \binits{S.D.}},
\bauthor{\bsnm{{Lin}}, \binits{R.P.}}:
\byear{2011},
\batitle{{Solar Source and Heliospheric Consequences of the 2010 April 3
  Coronal Mass Ejection: A Comprehensive View}}.
\bjtitle{\apj}
\bvolume{734},
\bfpage{84}.
\doiurl{10.1088/0004-637X/734/2/84}.
\adsurl{2011ApJ...734...84L}.
\end{barticle}
\endbibitem

\bibitem[\protect\citeauthoryear{{Lugaz} \textit{et~al.}}{2012}]{lugaz12}
\begin{barticle}
\bauthor{\bsnm{{Lugaz}}, \binits{N.}},
\bauthor{\bsnm{{Farrugia}}, \binits{C.J.}},
\bauthor{\bsnm{{Davies}}, \binits{J.A.}},
\bauthor{\bsnm{{M{\"o}stl}}, \binits{C.}},
\bauthor{\bsnm{{Davis}}, \binits{C.J.}},
\bauthor{\bsnm{{Roussev}}, \binits{I.I.}},
\bauthor{\bsnm{{Temmer}}, \binits{M.}}:
\byear{2012},
\batitle{{The Deflection of the Two Interacting Coronal Mass Ejections of 2010
  May 23-24 as Revealed by Combined in Situ Measurements and Heliospheric
  Imaging}}.
\bjtitle{\apj}
\bvolume{759},
\bfpage{68}.
\doiurl{10.1088/0004-637X/759/1/68}.
\adsurl{http://cdsads.u-strasbg.fr/abs/2012ApJ...759...68L}.
\end{barticle}
\endbibitem

\bibitem[\protect\citeauthoryear{{Lugaz} \textit{et~al.}}{2017}]{lugaz17}
\begin{barticle}
\bauthor{\bsnm{{Lugaz}}, \binits{N.}},
\bauthor{\bsnm{{Temmer}}, \binits{M.}},
\bauthor{\bsnm{{Wang}}, \binits{Y.}},
\bauthor{\bsnm{{Farrugia}}, \binits{C.J.}}:
\byear{2017},
\batitle{{The Interaction of Successive Coronal Mass Ejections: A Review}}.
\bjtitle{\solphys}
\bvolume{292},
\bfpage{64}.
\doiurl{10.1007/s11207-017-1091-6}.
\adsurl{2017SoPh..292...64L}.
\end{barticle}
\endbibitem

\bibitem[\protect\citeauthoryear{{Mason}, {Desai}, and {Li}}{2012}]{mason2012}
\begin{barticle}
\bauthor{\bsnm{{Mason}}, \binits{G.M.}},
\bauthor{\bsnm{{Desai}}, \binits{M.I.}},
\bauthor{\bsnm{{Li}}, \binits{G.}}:
\byear{2012},
\batitle{{Solar Cycle Abundance Variations in Corotating Interaction Regions:
  Evidence for a Suprathermal Ion Seed Population}}.
\bjtitle{\apjl}
\bvolume{748},
\bfpage{L31}.
\doiurl{10.1088/2041-8205/748/2/L31}.
\adsurl{2012ApJ...748L..31M}.
\end{barticle}
\endbibitem

\bibitem[\protect\citeauthoryear{{McComas} \textit{et~al.}}{1998}]{mccomas98}
\begin{barticle}
\bauthor{\bsnm{{McComas}}, \binits{D.J.}},
\bauthor{\bsnm{{Bame}}, \binits{S.J.}},
\bauthor{\bsnm{{Barker}}, \binits{P.}},
\bauthor{\bsnm{{Feldman}}, \binits{W.C.}},
\bauthor{\bsnm{{Phillips}}, \binits{J.L.}},
\bauthor{\bsnm{{Riley}}, \binits{P.}},
\bauthor{\bsnm{{Griffee}}, \binits{J.W.}}:
\byear{1998},
\batitle{{Solar Wind Electron Proton Alpha Monitor (SWEPAM) for the Advanced
  Composition Explorer}}.
\bjtitle{\ssr}
\bvolume{86},
\bfpage{563}.
\doiurl{10.1023/A:1005040232597}.
\adsurl{1998SSRv...86..563M}.
\end{barticle}
\endbibitem

\bibitem[\protect\citeauthoryear{{McNeice}, {Elliot}, and
  {Acebal}}{2011}]{mcneice11}
\begin{barticle}
\bauthor{\bsnm{{McNeice}}, \binits{P.}},
\bauthor{\bsnm{{Elliot}}, \binits{B.}},
\bauthor{\bsnm{{Acebal}}, \binits{A.}}:
\byear{2011},
\batitle{{Validation of community models}}.
\bjtitle{Space Weather}
\bvolume{9},
\bfpage{S10003}.
\doiurl{10.1029/2011SW000665}.
\end{barticle}
\endbibitem

\bibitem[\protect\citeauthoryear{{Moon} \textit{et~al.}}{2003}]{moon2003}
\begin{barticle}
\bauthor{\bsnm{{Moon}}, \binits{Y.-J.}},
\bauthor{\bsnm{{Choe}}, \binits{G.S.}},
\bauthor{\bsnm{{Wang}}, \binits{H.}},
\bauthor{\bsnm{{Park}}, \binits{Y.D.}}:
\byear{2003},
\batitle{{Sympathetic Coronal Mass Ejections}}.
\bjtitle{\apj}
\bvolume{588},
\bfpage{1176}.
\doiurl{10.1086/374270}.
\adsurl{2003ApJ...588.1176M}.
\end{barticle}
\endbibitem

\bibitem[\protect\citeauthoryear{{M{\"o}stl} \textit{et~al.}}{2010}]{mostl2010}
\begin{barticle}
\bauthor{\bsnm{{M{\"o}stl}}, \binits{C.}},
\bauthor{\bsnm{{Temmer}}, \binits{M.}},
\bauthor{\bsnm{{Rollett}}, \binits{T.}},
\bauthor{\bsnm{{Farrugia}}, \binits{C.J.}},
\bauthor{\bsnm{{Liu}}, \binits{Y.}},
\bauthor{\bsnm{{Veronig}}, \binits{A.M.}},
\bauthor{\bsnm{{Leitner}}, \binits{M.}},
\bauthor{\bsnm{{Galvin}}, \binits{A.B.}},
\bauthor{\bsnm{{Biernat}}, \binits{H.K.}}:
\byear{2010},
\batitle{{STEREO and Wind observations of a fast ICME flank triggering a
  prolonged geomagnetic storm on 5-7 April 2010}}.
\bjtitle{\grl}
\bvolume{37},
\bfpage{L24103}.
\doiurl{10.1029/2010GL045175}.
\adsurl{2010GeoRL..3724103M}.
\end{barticle}
\endbibitem

\bibitem[\protect\citeauthoryear{{M{\"o}stl} \textit{et~al.}}{2012}]{mostl12}
\begin{barticle}
\bauthor{\bsnm{{M{\"o}stl}}, \binits{C.}},
\bauthor{\bsnm{{Farrugia}}, \binits{C.J.}},
\bauthor{\bsnm{{Kilpua}}, \binits{E.K.J.}},
\bauthor{\bsnm{{Jian}}, \binits{L.K.}},
\bauthor{\bsnm{{Liu}}, \binits{Y.}},
\bauthor{\bsnm{{Eastwood}}, \binits{J.P.}},
\bauthor{\bsnm{{Harrison}}, \binits{R.A.}},
\bauthor{\bsnm{{Webb}}, \binits{D.F.}},
\bauthor{\bsnm{{Temmer}}, \binits{M.}},
\bauthor{\bsnm{{Odstrcil}}, \binits{D.}},
\bauthor{\bsnm{{Davies}}, \binits{J.A.}},
\bauthor{\bsnm{{Rollett}}, \binits{T.}},
\bauthor{\bsnm{{Luhmann}}, \binits{J.G.}},
\bauthor{\bsnm{{Nitta}}, \binits{N.}},
\bauthor{\bsnm{{Mulligan}}, \binits{T.}},
\bauthor{\bsnm{{Jensen}}, \binits{E.A.}},
\bauthor{\bsnm{{Forsyth}}, \binits{R.}},
\bauthor{\bsnm{{Lavraud}}, \binits{B.}},
\bauthor{\bsnm{{de Koning}}, \binits{C.A.}},
\bauthor{\bsnm{{Veronig}}, \binits{A.M.}},
\bauthor{\bsnm{{Galvin}}, \binits{A.B.}},
\bauthor{\bsnm{{Zhang}}, \binits{T.L.}},
\bauthor{\bsnm{{Anderson}}, \binits{B.J.}}:
\byear{2012},
\batitle{{Multi-point Shock and Flux Rope Analysis of Multiple Interplanetary
  Coronal Mass Ejections around 2010 August 1 in the Inner Heliosphere}}.
\bjtitle{The Astrophysical Journal}
\bvolume{758},
\bfpage{10}.
\doiurl{10.1088/0004-637X/758/1/10}.
\adsurl{2012ApJ...758...10M}.
\end{barticle}
\endbibitem

\bibitem[\protect\citeauthoryear{{Mrozek} \textit{et~al.}}{2013}]{mrozek13}
\begin{bchapter}
\bauthor{\bsnm{{Mrozek}}, \binits{T.}},
\bauthor{\bsnm{{Gburek}}, \binits{S.}},
\bauthor{\bsnm{{Siarkowski}}, \binits{M.}},
\bauthor{\bsnm{{Sylwester}}, \binits{B.}},
\bauthor{\bsnm{{Sylwester}}, \binits{J.}},
\bauthor{\bsnm{{K{\c e}pa}}, \binits{A.}},
\bauthor{\bsnm{{Gryciuk}}, \binits{M.}}:
\byear{2013},
\bctitle{{Solar flares observed simultaneously with SphinX, GOES and RHESSI}}.
In: \beditor{\bsnm{{Kosovichev}}, \binits{A.G.}},
\beditor{\bsnm{{de Gouveia Dal Pino}}, \binits{E.}},
\beditor{\bsnm{{Yan}}, \binits{Y.}} (eds.)
\bbtitle{Solar and Astrophysical Dynamos and Magnetic Activity},
\bsertitle{IAU Symposium}
\bseriesno{294},
\bfpage{571}.
\doiurl{10.1017/S1743921313003256}.
\adsurl{2013IAUS..294..571M}.
\end{bchapter}
\endbibitem

\bibitem[\protect\citeauthoryear{{Nolte} and {Roelof}}{1973}]{nolte73a}
\begin{barticle}
\bauthor{\bsnm{{Nolte}}, \binits{J.T.}},
\bauthor{\bsnm{{Roelof}}, \binits{E.C.}}:
\byear{1973},
\batitle{{Large-Scale Structure of the Interplanetary Medium, I: High Coronal
  Source Longitude of the Quiet-Time Solar Wind}}.
\bjtitle{Solar Physics}
\bvolume{33},
\bfpage{241}.
\doiurl{10.1007/BF00152395}.
\adsurl{1973SoPh...33..241N}.
\end{barticle}
\endbibitem

\bibitem[\protect\citeauthoryear{{Ogilvie} \textit{et~al.}}{1995}]{ogil1995}
\begin{barticle}
\bauthor{\bsnm{{Ogilvie}}, \binits{K.W.}},
\bauthor{\bsnm{{Chornay}}, \binits{D.J.}},
\bauthor{\bsnm{{Fritzenreiter}}, \binits{R.J.}},
\bauthor{\bsnm{{Hunsaker}}, \binits{F.}},
\bauthor{\bsnm{{Keller}}, \binits{J.}},
\bauthor{\bsnm{{Lobell}}, \binits{J.}},
\bauthor{\bsnm{{Miller}}, \binits{G.}},
\bauthor{\bsnm{{Scudder}}, \binits{J.D.}},
\bauthor{\bsnm{{Sittler}}, \binits{E.C.} \bsuffix{Jr.}},
\bauthor{\bsnm{{Torbert}}, \binits{R.B.}},
\bauthor{\bsnm{{Bodet}}, \binits{D.}},
\bauthor{\bsnm{{Needell}}, \binits{G.}},
\bauthor{\bsnm{{Lazarus}}, \binits{A.J.}},
\bauthor{\bsnm{{Steinberg}}, \binits{J.T.}},
\bauthor{\bsnm{{Tappan}}, \binits{J.H.}},
\bauthor{\bsnm{{Mavretic}}, \binits{A.}},
\bauthor{\bsnm{{Gergin}}, \binits{E.}}:
\byear{1995},
\batitle{{SWE, A Comprehensive Plasma Instrument for the Wind Spacecraft}}.
\bjtitle{\ssr}
\bvolume{71},
\bfpage{55}.
\doiurl{10.1007/BF00751326}.
\adsurl{1995SSRv...71...55O}.
\end{barticle}
\endbibitem

\bibitem[\protect\citeauthoryear{{Richardson} and {Cane}}{2004}]{rich04}
\begin{barticle}
\bauthor{\bsnm{{Richardson}}, \binits{I.G.}},
\bauthor{\bsnm{{Cane}}, \binits{H.V.}}:
\byear{2004},
\batitle{{Identification of interplanetary coronal mass ejections at 1 AU using
  multiple solar wind plasma composition anomalies}}.
\bjtitle{Journal of Geophysical Research (Space Physics)}
\bvolume{109},
\bfpage{A09104}.
\doiurl{10.1029/2004JA010598}.
\adsurl{2004JGRA..109.9104R}.
\end{barticle}
\endbibitem

\bibitem[\protect\citeauthoryear{{Richardson} and {Cane}}{2010}]{rich2010}
\begin{barticle}
\bauthor{\bsnm{{Richardson}}, \binits{I.G.}},
\bauthor{\bsnm{{Cane}}, \binits{H.V.}}:
\byear{2010},
\batitle{{Near-Earth Interplanetary Coronal Mass Ejections During Solar Cycle
  23 (1996 - 2009): Catalog and Summary of Properties}}.
\bjtitle{\solphys}
\bvolume{264},
\bfpage{189}.
\doiurl{10.1007/s11207-010-9568-6}.
\adsurl{2010SoPh..264..189R}.
\end{barticle}
\endbibitem

\bibitem[\protect\citeauthoryear{{Robbrecht}, {Patsourakos}, and
  {Vourlidas}}{2009}]{robb09}
\begin{barticle}
\bauthor{\bsnm{{Robbrecht}}, \binits{E.}},
\bauthor{\bsnm{{Patsourakos}}, \binits{S.}},
\bauthor{\bsnm{{Vourlidas}}, \binits{A.}}:
\byear{2009},
\batitle{{No Trace Left Behind: STEREO Observation of a Coronal Mass Ejection
  Without Low Coronal Signatures}}.
\bjtitle{\apj}
\bvolume{701},
\bfpage{283}.
\doiurl{10.1088/0004-637X/701/1/283}.
\adsurl{http://cdsads.u-strasbg.fr/abs/2009ApJ...701..283R}.
\end{barticle}
\endbibitem

\bibitem[\protect\citeauthoryear{{Rodkin}, {Shugay}, and
  {Slemzin}}{2016}]{rodkin16}
\begin{barticle}
\bauthor{\bsnm{{Rodkin}}, \binits{D.G.}},
\bauthor{\bsnm{{Shugay}}, \binits{Y.S.}},
\bauthor{\bsnm{{Slemzin}}, \binits{I.S.} \bsuffix{V.~A.and~{Veselovsky}}}:
\byear{2016},
\batitle{{Interaction of high-speed and transient fluxes of solar wind at the
  maximum of solar cycle 24}}.
\bjtitle{Bulletin of the Lebedev Physics Institute}
\bvolume{43},
\bfpage{287}.
\doiurl{10.3103/S1068335616090062}.
\end{barticle}
\endbibitem

\bibitem[\protect\citeauthoryear{{Rod'kin} \textit{et~al.}}{2016}]{rodkin2016a}
\begin{barticle}
\bauthor{\bsnm{{Rod'kin}}, \binits{D.G.}},
\bauthor{\bsnm{{Shugay}}, \binits{Y.S.}},
\bauthor{\bsnm{{Slemzin}}, \binits{V.A.}},
\bauthor{\bsnm{{Veselovskii}}, \binits{I.S.}}:
\byear{2016},
\batitle{{The effect of solar activity on the evolution of solar wind
  parameters during the rise of the 24th cycle}}.
\bjtitle{Solar System Research}
\bvolume{50},
\bfpage{44}.
\doiurl{10.1134/S0038094616010032}.
\adsurl{2016SoSyR..50...44R}.
\end{barticle}
\endbibitem

\bibitem[\protect\citeauthoryear{{Rodkin} \textit{et~al.}}{2017}]{rodkin2017}
\begin{barticle}
\bauthor{\bsnm{{Rodkin}}, \binits{D.}},
\bauthor{\bsnm{{Goryaev}}, \binits{F.}},
\bauthor{\bsnm{{Pagano}}, \binits{P.}},
\bauthor{\bsnm{{Gibb}}, \binits{G.}},
\bauthor{\bsnm{{Slemzin}}, \binits{V.}},
\bauthor{\bsnm{{Shugay}}, \binits{Y.}},
\bauthor{\bsnm{{Veselovsky}}, \binits{I.}},
\bauthor{\bsnm{{Mackay}}, \binits{D.H.}}:
\byear{2017},
\batitle{{Origin and Ion Charge State Evolution of Solar Wind Transients during
  4 - 7 August 2011}}.
\bjtitle{\solphys}
\bvolume{292},
\bfpage{90}.
\doiurl{10.1007/s11207-017-1109-0}.
\adsurl{2017SoPh..292...90R}.
\end{barticle}
\endbibitem

\bibitem[\protect\citeauthoryear{{Rouillard}
  \textit{et~al.}}{2010}]{rouillard2010}
\begin{barticle}
\bauthor{\bsnm{{Rouillard}}, \binits{A.P.}},
\bauthor{\bsnm{{Lavraud}}, \binits{B.}},
\bauthor{\bsnm{{Sheeley}}, \binits{N.R.}},
\bauthor{\bsnm{{Davies}}, \binits{J.A.}},
\bauthor{\bsnm{{Burlaga}}, \binits{L.F.}},
\bauthor{\bsnm{{Savani}}, \binits{N.P.}},
\bauthor{\bsnm{{Jacquey}}, \binits{C.}},
\bauthor{\bsnm{{Forsyth}}, \binits{R.J.}}:
\byear{2010},
\batitle{{White Light and In Situ Comparison of a Forming Merged Interaction
  Region}}.
\bjtitle{\apj}
\bvolume{719},
\bfpage{1385}.
\doiurl{10.1088/0004-637X/719/2/1385}.
\adsurl{2010ApJ...719.1385R}.
\end{barticle}
\endbibitem

\bibitem[\protect\citeauthoryear{{Schwenn} \textit{et~al.}}{2006}]{schwenn06}
\begin{barticle}
\bauthor{\bsnm{{Schwenn}}, \binits{R.}},
\bauthor{\bsnm{{Raymond}}, \binits{J.C.}},
\bauthor{\bsnm{{Alexander}}, \binits{D.}},
\bauthor{\bsnm{{Ciaravella}}, \binits{A.}},
\bauthor{\bsnm{{Gopalswamy}}, \binits{N.}},
\bauthor{\bsnm{{Howard}}, \binits{R.}},
\bauthor{\bsnm{{Hudson}}, \binits{H.}},
\bauthor{\bsnm{{Kaufmann}}, \binits{P.}},
\bauthor{\bsnm{{Klassen}}, \binits{A.}},
\bauthor{\bsnm{{Maia}}, \binits{D.}},
\bauthor{\bsnm{{Munoz-Martinez}}, \binits{G.}},
\bauthor{\bsnm{{Pick}}, \binits{M.}},
\bauthor{\bsnm{{Reiner}}, \binits{M.}},
\bauthor{\bsnm{{Srivastava}}, \binits{N.}},
\bauthor{\bsnm{{Tripathi}}, \binits{D.}},
\bauthor{\bsnm{{Vourlidas}}, \binits{A.}},
\bauthor{\bsnm{{Wang}}, \binits{Y.-M.}},
\bauthor{\bsnm{{Zhang}}, \binits{J.}}:
\byear{2006},
\batitle{{Coronal Observations of CMEs. Report of Working Group A}}.
\bjtitle{Space Science Reviews}
\bvolume{123},
\bfpage{127}.
\doiurl{10.1007/s11214-006-9016-y}.
\adsurl{2006SSRv..123..127S}.
\end{barticle}
\endbibitem

\bibitem[\protect\citeauthoryear{{Shen} \textit{et~al.}}{2017}]{shen2017}
\begin{barticle}
\bauthor{\bsnm{{Shen}}, \binits{F.}},
\bauthor{\bsnm{{Wang}}, \binits{Y.}},
\bauthor{\bsnm{{Shen}}, \binits{C.}},
\bauthor{\bsnm{{Feng}}, \binits{X.}}:
\byear{2017},
\batitle{{On the Collision Nature of Two Coronal Mass Ejections: A Review}}.
\bjtitle{\solphys}
\bvolume{292},
\bfpage{104}.
\doiurl{10.1007/s11207-017-1129-9}.
\adsurl{2017SoPh..292..104S}.
\end{barticle}
\endbibitem

\bibitem[\protect\citeauthoryear{{Shi} \textit{et~al.}}{2015}]{shi15}
\begin{barticle}
\bauthor{\bsnm{{Shi}}, \binits{T.}},
\bauthor{\bsnm{{Wang}}, \binits{Y.}},
\bauthor{\bsnm{{Wan}}, \binits{L.}},
\bauthor{\bsnm{{Cheng}}, \binits{X.}},
\bauthor{\bsnm{{Ding}}, \binits{M.}},
\bauthor{\bsnm{{Zhang}}, \binits{J.}}:
\byear{2015},
\batitle{{Predicting the Arrival Time of Coronal Mass Ejections with the
  Graduated Cylindrical Shell and Drag Force Model}}.
\bjtitle{The Astrophysical Journal}
\bvolume{806},
\bfpage{271}.
\doiurl{10.1088/0004-637X/806/2/271}.
\adsurl{http://cdsads.u-strasbg.fr/abs/2015ApJ...806..271S}.
\end{barticle}
\endbibitem

\bibitem[\protect\citeauthoryear{{Shugay} \textit{et~al.}}{2011}]{shugay2011}
\begin{barticle}
\bauthor{\bsnm{{Shugay}}, \binits{Y.S.}},
\bauthor{\bsnm{{Veselovsky}}, \binits{I.S.}},
\bauthor{\bsnm{{Seaton}}, \binits{D.B.}},
\bauthor{\bsnm{{Berghmans}}, \binits{D.}}:
\byear{2011},
\batitle{{Hierarchical approach to forecasting recurrent solar wind streams}}.
\bjtitle{Solar System Research}
\bvolume{45},
\bfpage{546}.
\doiurl{10.1134/S0038094611060086}.
\adsurl{2011SoSyR..45..546S}.
\end{barticle}
\endbibitem

\bibitem[\protect\citeauthoryear{{Shugay} \textit{et~al.}}{2017}]{shugay17}
\begin{barticle}
\bauthor{\bsnm{{Shugay}}, \binits{Y.S.}},
\bauthor{\bsnm{{Veselovsky}}, \binits{I.S.}},
\bauthor{\bsnm{{Slemzin}}, \binits{V.A.}},
\bauthor{\bsnm{{Yermolaev}}, \binits{Y.I.}},
\bauthor{\bsnm{{Rodkin}}, \binits{D.G.}}:
\byear{2017},
\batitle{{Possible causes of the discrepancy between the predicted and observed
  parameters of high-speed solar wind streams}}.
\bjtitle{Cosmic Research}
\bvolume{55},
\bfpage{20}.
\doiurl{10.1134/S0010952517010087}.
\adsurl{2017CosRe..55...20S}.
\end{barticle}
\endbibitem

\bibitem[\protect\citeauthoryear{{Smith} \textit{et~al.}}{1998}]{smith1998}
\begin{barticle}
\bauthor{\bsnm{{Smith}}, \binits{C.W.}},
\bauthor{\bsnm{{L'Heureux}}, \binits{J.}},
\bauthor{\bsnm{{Ness}}, \binits{N.F.}},
\bauthor{\bsnm{{Acu{\~n}a}}, \binits{M.H.}},
\bauthor{\bsnm{{Burlaga}}, \binits{L.F.}},
\bauthor{\bsnm{{Scheifele}}, \binits{J.}}:
\byear{1998},
\batitle{{The ACE Magnetic Fields Experiment}}.
\bjtitle{\ssr}
\bvolume{86},
\bfpage{613}.
\doiurl{10.1023/A:1005092216668}.
\adsurl{1998SSRv...86..613S}.
\end{barticle}
\endbibitem

\bibitem[\protect\citeauthoryear{{Somov}}{2013}]{somov2013}
\begin{bbook}
\beditor{\bsnm{{Somov}}, \binits{B.V.}} (ed.):
\byear{2013},
\bbtitle{{Plasma Astrophysics, Part II}},
\bsertitle{Astrophysics and Space Science Library}
\bseriesno{392}.
\doiurl{10.1007/978-1-4614-4295-0}.
\adsurl{2013ASSL..392.....S}.
\end{bbook}
\endbibitem

\bibitem[\protect\citeauthoryear{{Stone} \textit{et~al.}}{1998}]{stone98}
\begin{barticle}
\bauthor{\bsnm{{Stone}}, \binits{E.C.}},
\bauthor{\bsnm{{Frandsen}}, \binits{A.M.}},
\bauthor{\bsnm{{Mewaldt}}, \binits{R.A.}},
\bauthor{\bsnm{{Christian}}, \binits{E.R.}},
\bauthor{\bsnm{{Margolies}}, \binits{D.}},
\bauthor{\bsnm{{Ormes}}, \binits{J.F.}},
\bauthor{\bsnm{{Snow}}, \binits{F.}}:
\byear{1998},
\batitle{{The Advanced Composition Explorer}}.
\bjtitle{\ssr}
\bvolume{86},
\bfpage{1}.
\doiurl{10.1023/A:1005082526237}.
\adsurl{http://cdsads.u-strasbg.fr/abs/1998SSRv...86....1S}.
\end{barticle}
\endbibitem

\bibitem[\protect\citeauthoryear{{Temmer} \textit{et~al.}}{2011}]{temmer2011}
\begin{barticle}
\bauthor{\bsnm{{Temmer}}, \binits{M.}},
\bauthor{\bsnm{{Rollett}}, \binits{T.}},
\bauthor{\bsnm{{M{\"o}stl}}, \binits{C.}},
\bauthor{\bsnm{{Veronig}}, \binits{A.M.}},
\bauthor{\bsnm{{Vr{\v s}nak}}, \binits{B.}},
\bauthor{\bsnm{{Odstr{\v c}il}}, \binits{D.}}:
\byear{2011},
\batitle{{Influence of the Ambient Solar Wind Flow on the Propagation Behavior
  of Interplanetary Coronal Mass Ejections}}.
\bjtitle{\apj}
\bvolume{743},
\bfpage{101}.
\doiurl{10.1088/0004-637X/743/2/101}.
\adsurl{2011ApJ...743..101T}.
\end{barticle}
\endbibitem

\bibitem[\protect\citeauthoryear{{Temmer} \textit{et~al.}}{2012}]{temmer12}
\begin{barticle}
\bauthor{\bsnm{{Temmer}}, \binits{M.}},
\bauthor{\bsnm{{Vr{\v s}nak}}, \binits{B.}},
\bauthor{\bsnm{{Rollett}}, \binits{T.}},
\bauthor{\bsnm{{Bein}}, \binits{B.}},
\bauthor{\bsnm{{de Koning}}, \binits{C.A.}},
\bauthor{\bsnm{{Liu}}, \binits{Y.}},
\bauthor{\bsnm{{Bosman}}, \binits{E.}},
\bauthor{\bsnm{{Davies}}, \binits{J.A.}},
\bauthor{\bsnm{{M{\"o}stl}}, \binits{C.}},
\bauthor{\bsnm{{{\v Z}ic}}, \binits{T.}},
\bauthor{\bsnm{{Veronig}}, \binits{A.M.}},
\bauthor{\bsnm{{Bothmer}}, \binits{V.}},
\bauthor{\bsnm{{Harrison}}, \binits{R.}},
\bauthor{\bsnm{{Nitta}}, \binits{N.}},
\bauthor{\bsnm{{Bisi}}, \binits{M.}},
\bauthor{\bsnm{{Flor}}, \binits{O.}},
\bauthor{\bsnm{{Eastwood}}, \binits{J.}},
\bauthor{\bsnm{{Odstrcil}}, \binits{D.}},
\bauthor{\bsnm{{Forsyth}}, \binits{R.}}:
\byear{2012},
\batitle{{Characteristics of Kinematics of a Coronal Mass Ejection during the
  2010 August 1 CME-CME Interaction Event}}.
\bjtitle{The Astrophysical Journal}
\bvolume{749},
\bfpage{57}.
\doiurl{10.1088/0004-637X/749/1/57}.
\adsurl{2012ApJ...749...57T}.
\end{barticle}
\endbibitem

\bibitem[\protect\citeauthoryear{{Temmer} \textit{et~al.}}{2017}]{temmer2017}
\begin{barticle}
\bauthor{\bsnm{{Temmer}}, \binits{M.}},
\bauthor{\bsnm{{Reiss}}, \binits{M.A.}},
\bauthor{\bsnm{{Nikolic}}, \binits{L.}},
\bauthor{\bsnm{{Hofmeister}}, \binits{S.J.}},
\bauthor{\bsnm{{Veronig}}, \binits{A.M.}}:
\byear{2017},
\batitle{{Preconditioning of Interplanetary Space Due to Transient CME
  Disturbances}}.
\bjtitle{\apj}
\bvolume{835},
\bfpage{141}.
\doiurl{10.3847/1538-4357/835/2/141}.
\adsurl{2017ApJ...835..141T}.
\end{barticle}
\endbibitem

\bibitem[\protect\citeauthoryear{{{\v Z}ic}, {Vr{\v s}nak}, and
  {Temmer}}{2015}]{zic15}
\begin{barticle}
\bauthor{\bsnm{{{\v Z}ic}}, \binits{T.}},
\bauthor{\bsnm{{Vr{\v s}nak}}, \binits{B.}},
\bauthor{\bsnm{{Temmer}}, \binits{M.}}:
\byear{2015},
\batitle{{Heliospheric Propagation of Coronal Mass Ejections: Drag-based Model
  Fitting}}.
\bjtitle{The Astrophysical Journal Supplement Series}
\bvolume{218},
\bfpage{32}.
\doiurl{10.1088/0067-0049/218/2/32}.
\adsurl{2015ApJS..218...32Z}.
\end{barticle}
\endbibitem

\bibitem[\protect\citeauthoryear{{Verbanac} \textit{et~al.}}{2013}]{verbanac13}
\begin{barticle}
\bauthor{\bsnm{{Verbanac}}, \binits{G.-}},
\bauthor{\bsnm{{{\v Z}ivkovi{\'c}}}, \binits{S.-}},
\bauthor{\bsnm{{Vr{\v s}nak}}, \binits{B.-}},
\bauthor{\bsnm{{Bandi{\'c}}}, \binits{M.-}},
\bauthor{\bsnm{{Hojsak}}, \binits{T.-}}:
\byear{2013},
\batitle{{Comparison of geoeffectiveness of coronal mass ejections and
  corotating interaction regions}}.
\bjtitle{Astronomy \& Astrophysics}
\bvolume{558},
\bfpage{A85}.
\doiurl{10-1051/0004-6361/201220417}.
\adsurl{http://cdsads-u-strasbg-fr/abs/2013A\%26A---558A--85V}.
\end{barticle}
\endbibitem

\bibitem[\protect\citeauthoryear{{von Steiger}
  \textit{et~al.}}{1992}]{steiger92}
\begin{barticle}
\bauthor{\bsnm{{von Steiger}}, \binits{R.}},
\bauthor{\bsnm{{Christon}}, \binits{S.P.}},
\bauthor{\bsnm{{Gloeckler}}, \binits{G.}},
\bauthor{\bsnm{{Ipavich}}, \binits{F.M.}}:
\byear{1992},
\batitle{{Variable carbon and oxygen abundances in the solar wind as observed
  in earth's magnetosheath by AMPTE/CCE}}.
\bjtitle{\apj}
\bvolume{389},
\bfpage{791}.
\doiurl{10.1086/171252}.
\adsurl{1992ApJ...389..791V}.
\end{barticle}
\endbibitem

\bibitem[\protect\citeauthoryear{{Webb} and {Howard}}{2012}]{fwebb12}
\begin{barticle}
\bauthor{\bsnm{{Webb}}, \binits{D.F.}},
\bauthor{\bsnm{{Howard}}, \binits{T.A.}}:
\byear{2012},
\batitle{{Coronal Mass Ejections: Observations}}.
\bjtitle{Living Reviews in Solar Physics}
\bvolume{9}.
\doiurl{10.12942/lrsp-2012-3}.
\adsurl{2012LRSP....9....3W}.
\end{barticle}
\endbibitem

\bibitem[\protect\citeauthoryear{{Wu} \textit{et~al.}}{2016}]{wu16}
\begin{barticle}
\bauthor{\bsnm{{Wu}}, \binits{C.-C.}},
\bauthor{\bsnm{{Liou}}, \binits{K.}},
\bauthor{\bsnm{{Lepping}}, \binits{R.P.}},
\bauthor{\bsnm{{Hutting}}, \binits{L.}},
\bauthor{\bsnm{{Plunkett}}, \binits{S.}},
\bauthor{\bsnm{{Howard}}, \binits{R.A.}},
\bauthor{\bsnm{{Socker}}, \binits{D.}}:
\byear{2016},
\batitle{{The first super geomagnetic storm of solar cycle 24: ``The St.
  Patrick's day event (17 March 2015)''}}.
\bjtitle{Earth, Planets, and Space}
\bvolume{68},
\bfpage{151}.
\doiurl{10.1186/s40623-016-0525-y}.
\adsurl{2016EP\%26S...68..151W}.
\end{barticle}
\endbibitem

\bibitem[\protect\citeauthoryear{{Yashiro} and {Gopalswamy}}{2009}]{yashiro09}
\begin{bchapter}
\bauthor{\bsnm{{Yashiro}}, \binits{S.-}},
\bauthor{\bsnm{{Gopalswamy}}, \binits{N.-}}:
\byear{2009},
\bctitle{{Statistical relationship between solar flares and coronal mass
  ejections}}.
In: \beditor{\bsnm{{Gopalswamy}}, \binits{N.-}},
\beditor{\bsnm{{Webb}}, \binits{D.-F.-}} (eds.)
\bbtitle{Universal Heliophysical Processes},
\bsertitle{IAU Symposium}
\bseriesno{257},
\bfpage{233}.
\doiurl{10-1017/S1743921309029342}.
\adsurl{http://cdsads-u-strasbg-fr/abs/2009IAUS--257--233Y}.
\end{bchapter}
\endbibitem

\bibitem[\protect\citeauthoryear{{Yermolaev}
  \textit{et~al.}}{2007}]{yermolaev2007}
\begin{barticle}
\bauthor{\bsnm{{Yermolaev}}, \binits{Y.I.}},
\bauthor{\bsnm{{Yermolaev}}, \binits{M.Y.}},
\bauthor{\bsnm{{Lodkina}}, \binits{I.G.}},
\bauthor{\bsnm{{Nikolaeva}}, \binits{N.S.}}:
\byear{2007},
\batitle{{Statistical investigation of heliospheric conditions resulting in
  magnetic storms}}.
\bjtitle{Cosmic Research}
\bvolume{45},
\bfpage{1}.
\doiurl{10.1134/S0010952507010017}.
\adsurl{2007CosRe..45....1Y}.
\end{barticle}
\endbibitem

\bibitem[\protect\citeauthoryear{{Yermolaev}
  \textit{et~al.}}{2012}]{yermolaev12}
\begin{barticle}
\bauthor{\bsnm{{Yermolaev}}, \binits{Y.-I.-}},
\bauthor{\bsnm{{Nikolaeva}}, \binits{N.-S.-}},
\bauthor{\bsnm{{Lodkina}}, \binits{I.-G.-}},
\bauthor{\bsnm{{Yermolaev}}, \binits{M.-Y.-}}:
\byear{2012},
\batitle{{Geoeffectiveness and efficiency of CIR, sheath, and ICME in
  generation of magnetic storms}}.
\bjtitle{Journal of Geophysical Research (Space Physics)}
\bvolume{117},
\bfpage{A00L07}.
\doiurl{10-1029/2011JA017139}.
\adsurl{http://cdsads-u-strasbg-fr/abs/2012JGRA--117-0L07Y}.
\end{barticle}
\endbibitem

\bibitem[\protect\citeauthoryear{{Zhang} and {Burlaga}}{1988}]{zhang1988}
\begin{barticle}
\bauthor{\bsnm{{Zhang}}, \binits{G.}},
\bauthor{\bsnm{{Burlaga}}, \binits{L.F.}}:
\byear{1988},
\batitle{{Magnetic clouds, geomagnetic disturbances, and cosmic ray
  decreases}}.
\bjtitle{\jgr}
\bvolume{93},
\bfpage{2511}.
\doiurl{10.1029/JA093iA04p02511}.
\adsurl{1988JGR....93.2511Z}.
\end{barticle}
\endbibitem

\bibitem[\protect\citeauthoryear{{Zhang} and {Wang}}{2002}]{zhang2002}
\begin{barticle}
\bauthor{\bsnm{{Zhang}}, \binits{J.}},
\bauthor{\bsnm{{Wang}}, \binits{J.}}:
\byear{2002},
\batitle{{Are Homologous Flare-Coronal Mass Ejection Events Triggered by Moving
  Magnetic Features?}}
\bjtitle{\apjl}
\bvolume{566},
\bfpage{L117}.
\doiurl{10.1086/339660}.
\adsurl{2002ApJ...566L.117Z}.
\end{barticle}
\endbibitem

\bibitem[\protect\citeauthoryear{{Zhang} \textit{et~al.}}{2004}]{zhang2004}
\begin{barticle}
\bauthor{\bsnm{{Zhang}}, \binits{J.}},
\bauthor{\bsnm{{Liemohn}}, \binits{M.W.}},
\bauthor{\bsnm{{Kozyra}}, \binits{J.U.}},
\bauthor{\bsnm{{Lynch}}, \binits{B.J.}},
\bauthor{\bsnm{{Zurbuchen}}, \binits{T.H.}}:
\byear{2004},
\batitle{{A statistical study of the geoeffectiveness of magnetic clouds during
  high solar activity years}}.
\bjtitle{Journal of Geophysical Research (Space Physics)}
\bvolume{109},
\bfpage{A09101}.
\doiurl{10.1029/2004JA010410}.
\adsurl{2004JGRA..109.9101Z}.
\end{barticle}
\endbibitem

\bibitem[\protect\citeauthoryear{{Zhang} \textit{et~al.}}{2007}]{zhang07}
\begin{barticle}
\bauthor{\bsnm{{Zhang}}, \binits{J.}},
\bauthor{\bsnm{{Richardson}}, \binits{I.G.}},
\bauthor{\bsnm{{Webb}}, \binits{D.F.}},
\bauthor{\bsnm{{Gopalswamy}}, \binits{N.}},
\bauthor{\bsnm{{Huttunen}}, \binits{E.}},
\bauthor{\bsnm{{Kasper}}, \binits{J.C.}},
\bauthor{\bsnm{{Nitta}}, \binits{N.V.}},
\bauthor{\bsnm{{Poomvises}}, \binits{W.}},
\bauthor{\bsnm{{Thompson}}, \binits{B.J.}},
\bauthor{\bsnm{{Wu}}, \binits{C.-C.}},
\bauthor{\bsnm{{Yashiro}}, \binits{S.}},
\bauthor{\bsnm{{Zhukov}}, \binits{A.N.}}:
\byear{2007},
\batitle{{Solar and interplanetary sources of major geomagnetic storms (Dst $<
  = -100 nT$) during 1996-2005}}.
\bjtitle{Journal of Geophysical Research (Space Physics)}
\bvolume{112},
\bfpage{A10102}.
\doiurl{10.1029/2007JA012321}.
\adsurl{2007JGRA..11210102Z}.
\end{barticle}
\endbibitem

\bibitem[\protect\citeauthoryear{{Zhao} \textit{et~al.}}{2016}]{zhao16}
\begin{bchapter}
\bauthor{\bsnm{{Zhao}}, \binits{L.}},
\bauthor{\bsnm{{Landi}}, \binits{E.}},
\bauthor{\bsnm{{Kocher}}, \binits{M.}},
\bauthor{\bsnm{{Lepri}}, \binits{S.T.}},
\bauthor{\bsnm{{Fisk}}, \binits{L.A.}}:
\byear{2016},
\bctitle{{Anomalously low C6+/C5+ ratio in solar wind: ACE/SWICS observation}}.
In: \bbtitle{AIP Publishing LLC},
\bsertitle{AIP Conference Proceedings}
\bseriesno{1720},
\bfpage{020006}.
\doiurl{10.1063/1.4943807}.
\end{bchapter}
\endbibitem

\bibitem[\protect\citeauthoryear{{Zhukov}}{2007}]{zhukov2007}
\begin{bchapter}
\bauthor{\bsnm{{Zhukov}}, \binits{A.N.}}:
\byear{2007},
\bctitle{{Using CME Observations for Geomagnetic Storm Forecasting}}.
In: \beditor{\bsnm{{Lilensten}}, \binits{J.}} (ed.)
\bbtitle{Space Weather : Research Towards Applications in Europe 2nd European
  Space Weather Week (ESWW2)},
\bsertitle{Astrophysics and Space Science Library}
\bseriesno{344},
\bfpage{5}.
\doiurl{10.1007/1-4020-5446-7_2}.
\adsurl{2007ASSL..344....5Z}.
\end{bchapter}
\endbibitem

\bibitem[\protect\citeauthoryear{{Zurbuchen} and
  {Richardson}}{2006}]{zurbuchen06}
\begin{barticle}
\bauthor{\bsnm{{Zurbuchen}}, \binits{T.H.}},
\bauthor{\bsnm{{Richardson}}, \binits{I.G.}}:
\byear{2006},
\batitle{{In-Situ Solar Wind and Magnetic Field Signatures of Interplanetary
  Coronal Mass Ejections}}.
\bjtitle{Space Science Reviews}
\bvolume{123},
\bfpage{31}.
\doiurl{10.1007/s11214-006-9010-4}.
\adsurl{http://cdsads.u-strasbg.fr/abs/2006SSRv..123...31Z}.
\end{barticle}
\endbibitem

\bibitem[\protect\citeauthoryear{{Zurbuchen}
  \textit{et~al.}}{1999}]{zurbuch1999}
\begin{barticle}
\bauthor{\bsnm{{Zurbuchen}}, \binits{T.H.}},
\bauthor{\bsnm{{Hefti}}, \binits{S.}},
\bauthor{\bsnm{{Fisk}}, \binits{L.A.}},
\bauthor{\bsnm{{Gloeckler}}, \binits{G.}},
\bauthor{\bsnm{{von Steiger}}, \binits{R.}}:
\byear{1999},
\batitle{{The Transition Between Fast and Slow Solar Wind from Composition
  Data}}.
\bjtitle{\ssr}
\bvolume{87},
\bfpage{353}.
\doiurl{10.1023/A:1005126718714}.
\adsurl{1999SSRv...87..353Z}.
\end{barticle}
\endbibitem

\end{thebibliography}

\end{article}

\end{document}